\documentclass[pre,twocolumn,twoside,byrevtex,superscriptaddress,floatfix]{revtex4-1}

\usepackage{graphicx,epsfig,verbatim,enumerate}
\usepackage{amssymb,amsmath}
\usepackage{ifthen}

\newboolean{twocolswitch}
\newboolean{normalrefs}

\usepackage{color}

\definecolor{lightgrey}{rgb}{0.5,0.5,0.5}

\usepackage{array}
\usepackage{longtable}
\usepackage{rotating}

\newcommand{\sindex}[1]{}
\newcommand{\nindex}[1]{}

\newcommand{\www}[1]{\url{#1}}

\newcommand{\PreserveBackslash}[1]{\let\temp=\\#1\let\\=\temp}
\newcommand{\PBS}[1]{\let\temp=\\#1\let\\=\temp}

\newcommand{\degree}{\, \circ}
\newcommand{\tempf}[1]{${#1}^{\degree}F$}

\newcommand{\tempfc}[2]{${#1}^{\degree}F$ (${#2}^{\degree}C$)}

\newcommand{\plainlatexonly}[1]{}

\setboolean{twocolswitch}{true}

\begin{document}

\title{\protect
Tracking climate change through the spatiotemporal dynamics 
of the Teletherms, the statistically hottest and coldest days of the year
}

\author{
\firstname{Peter Sheridan}
\surname{Dodds}
}
\email{peter.dodds@uvm.edu}

\affiliation{
  Computational Story Lab,
  Department of Mathematics and Statistics,
  Vermont Complex Systems Center,
  \&
  the Vermont Advanced Computing Center,
  University of Vermont,
  Burlington,
  VT, 05401
}

\author{
\firstname{Lewis}
\surname{Mitchell}
}

\email{lewis.mitchell@adelaide.edu.au}

\affiliation{
  School of Mathematical Sciences,
  North Terrace Campus,
  The University of Adelaide,
  SA 5005, Australia
}

\author{
\firstname{Andrew J.}
\surname{Reagan}
}
\email{andrew.reagan@uvm.edu}

\affiliation{
  Computational Story Lab,
  Department of Mathematics and Statistics,
  Vermont Complex Systems Center,
  \&
  the Vermont Advanced Computing Center,
  University of Vermont,
  Burlington,
  VT, 05401
}

\author{
\firstname{Christopher M.}
\surname{Danforth}
}
\email{chris.danforth@uvm.edu}

\affiliation{
  Computational Story Lab,
  Department of Mathematics and Statistics,
  Vermont Complex Systems Center,
  \&
  the Vermont Advanced Computing Center,
  University of Vermont,
  Burlington,
  VT, 05401
}

\date{\today}

\begin{abstract}
  \protect
  Instabilities and long term shifts in seasons, whether induced
by natural drivers or human activities, pose great disruptive 
threats to ecological, agricultural, and social systems.
Here, we propose, measure, and explore two fundamental
markers of location-sensitive seasonal variations: the Summer and Winter Teletherms---the on-average annual
dates of the hottest and coldest days of the year. 
We analyse daily temperature extremes recorded 
at 1218 stations
across the contiguous United States 
from 1853--2012,
and observe large regional variation with the Summer
Teletherm falling up to 90 days after the Summer Solstice,
and 50 days for the Winter Teletherm after the Winter Solstice.
We show that Teletherm temporal dynamics are substantive with 
clear and in some cases dramatic shifts reflective of 
system bifurcations.
We also compare recorded daily temperature extremes
with output from two regional climate models
finding considerable though relatively unbiased error.
Our work demonstrates that Teletherms are an intuitive, powerful, and statistically sound 
measure of local climate change, and that they pose detailed, stringent
challenges for future theoretical and computational models.

\end{abstract}

\maketitle

\textbf{Logline:}
This paper introduces, formalizes, and explores two fundamental
climatological and seasonal markers: the Summer and Winter
Teletherms---the on-average hottest and coldest days of the year.
Across the contiguous United States,
the variation of the Teletherms---in date, extent, and
temperature---is found  to be highly variable spatiotemporally
with local coherence.
The Teletherms reveal complex climate change histories over many
scales, including bifurcations and instabilities, and provide
stringent, detailed challenges to models and theory.


\section*{Introduction}

Day length and temperature are two 
of the most important driving factors
for life on Earth and for human culture.
While evidently strongly coupled,
their relationship is not a simple one in detail.

Due to the regularity of celestial and planetary motion
and the relative ease with which sun position can
be recorded, the Solstices and Equinoxes have been
determined and commemorated by cultures around the world
for thousands of years (e.g., Stonehenge), long before being scientifically understood.
We thus know with great precision when the longest and shortest day of the year will
be, but what about the on-average hottest and coldest days?

Temperature behaves stochastically
with highs and lows on a specific date 
potentially differing greatly relative
to surrounding dates and across years.
Compounding temperature's unevenness
is that reliable measurement has only been
realized in the last few hundred years.
Indeed, widespread, systematic recording in the United States,
which we study here, only began in the late 1800s.
We are only now in a position to capitalize on sufficiently large data sets
to give a reasonably solid answer to our question.

We propose to call the dates of on-average extreme temperature the \textit{Teletherms},
using the Greek roots \textit{tele} for distant and \textit{therm} for heat.
This construction is patterned after the Latin origin of Solstice 
with \textit{sol} for sun and \textit{stit} for stationary.

As we will find, 
the Teletherms and their temperatures are not fixed but vary in both space and time.
In particular, we will show that across the United States, the dynamics of the
Teletherms are locally coherent but overall highly variable, revealing intricate patterns 
including bifurcations in dates and both warming and cooling.
For many regions, we will also demonstrate that the Teletherm is more appropriately
acknowledged as occurring over a range of dates rather than a single one.
We will therefore also speak both of each location's 
single day Teletherm and its \textit{Teletherm Period} 
which we define below in a pragmatic fashion.

Our conception of the Teletherm is related to but differs from existing
meteorological quantities drawn from stations around the United States.
The National Oceanic and Atmospheric Administration (NOAA) captures
`climate normals': 30 year averages at day, month,
season, and year resolutions for a range
of quantities including mean, maximum, and minimum temperatures;
precipitation; and snowfall~\cite{arguez2012a}.
Climate normals are made available to and used broadly by the public.
For example, monthly averages are of great use to people traveling to new areas.
NOAA and the National Weather Service provide
the Local Climate Analysis Tool (LCAT) at \url{http://nws.weather.gov/lcat/home}
for people to explore historical and recent climate dynamics.
As we explain below, we estimate the Teletherms'
aspects---date, temperature, extent, and period---from a daily maximum and
minimum temperature data set, and as such our contribution can
be seen as building a new lens for the United States' rich meteorological data set.
We accompany our paper with an interactive site at
\url{http://panometer.org/instruments/teletherms}
to enable those interested to examine climate dynamics
through the Teletherms.
If the notion of the Teletherm becomes standard, we would hope
a version of this site might eventually be incorporated into the LCAT.

Despite the evident imperative of quantifying climate change,
the task has proven to be both scientifically complex~\cite{stainforth2005a,karl2015a,barkemeyer2015a,oreilly2015a}
and politically fraught and controversial~\cite{antilla2005a,lempert2015a}.
Tied as they are to the changing of the seasons~\cite{mann1996a,sparks2002a,schwartz2006a,stine2009a,betts2011a},
Teletherm dynamics matter for ecological stability,
agriculture, the Earth's water cycle, 
the livability of cities~\cite{lai2016a},
and cultural and religious observances.

By formalizing these annual turning points in temperature we hope to help advance our
collective understanding of and ability to discern climate change.
While we will make a number of general observations regarding Teletherm
dynamics, the central objective of our present work
is the introduction of a statistically sound quantification
of these two fundamental aspects of the annual climate cycle,
with the hope of both expanding and challenging future work on climate dynamics.

We structure our paper as follows.
We first make some basic observations about the 
historical weather data set which we build our analysis around,
along with a few details about our approach.
We then present our main findings,
describing and testing our approach to determining 
Teletherms and Teletherm Periods at specific locations,
highlighting a few of the extreme locations such
as the hottest Summer Teletherm and coldest Winter Teletherm.
Moving out from individual stations, we then 
examine a range of results 
for the contiguous United States.
We first show that the Winter and Summer Teletherms
vary strongly according to geographic location.
We then explore the temporal dynamics of regional Teletherms, and discuss
their relationship to climate change.
Finally, we compare empirical Teletherm dates with 
those produced by two Regional Climate Models (RCMs).
To close, we put forward a few concluding remarks, contemplating
future directions.

We provide a complete set of figures and code
as part of the paper's online appendices 
at 
\url{http://compstorylab.org/share/papers/dodds2015c}.

\section*{Data}
\label{sec:teletherm.data-and-analysis}
      
\begin{figure}[tbp!]
  \centering
\includegraphics[width=\columnwidth]{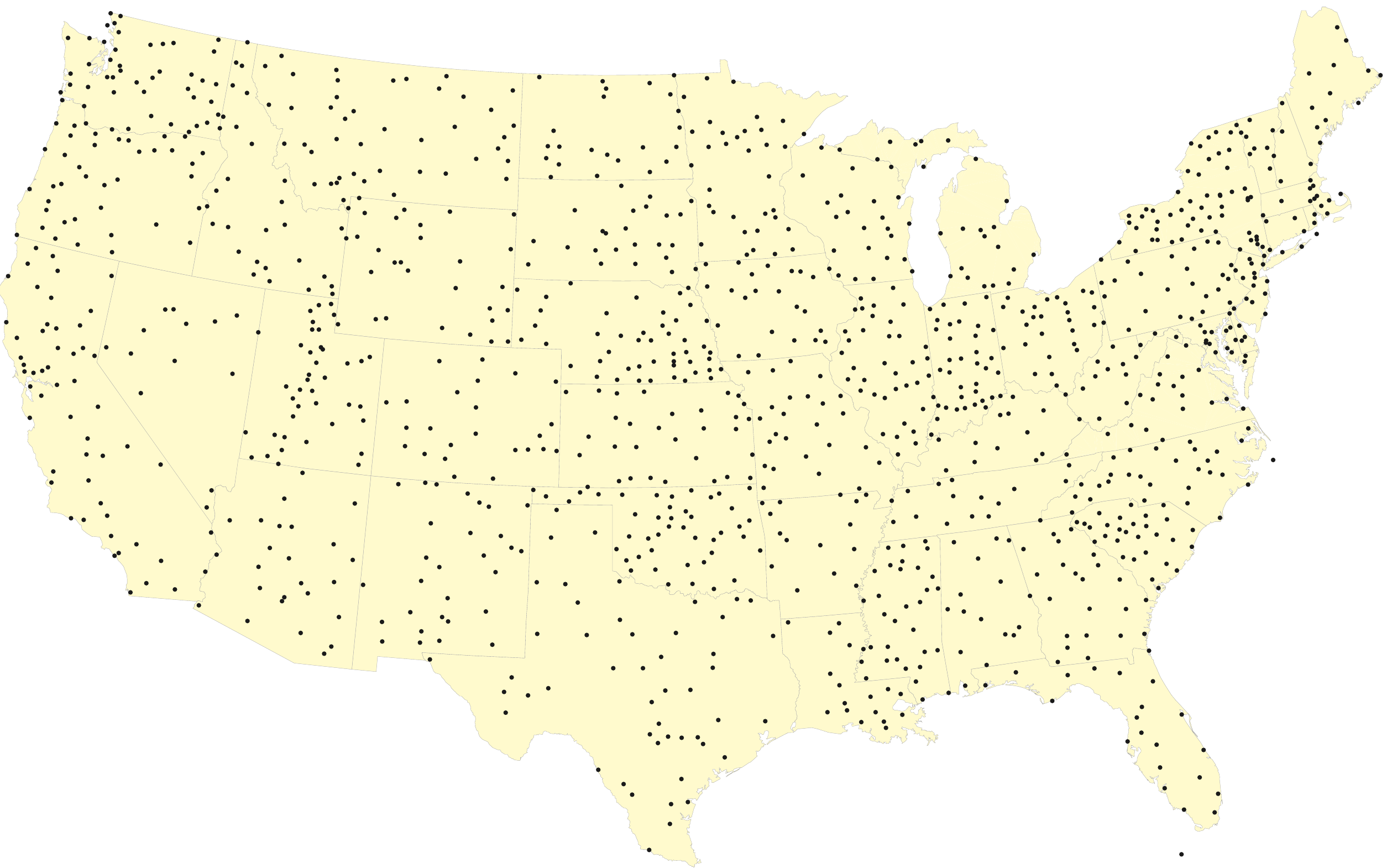}
  \caption{
    Locations of 1218 weather stations 
    represented  in the United States Historical Climatology Network
    (USCHN) data set (version 2.5 through 2012)~\cite{menne2012a}.
    The distribution indicates relatively uniform coverage of the 48 contiguous states.
  }
  \label{fig:teletherm_map002}
\end{figure}

We consider daily records of maximum and minimum temperatures
for 1218 stations distributed across the
contiguous United States for the time period
1853--2012.
We draw on the United States Historical Climatology Network (USCHN)
data set (version 2.5 through 2012)~\cite{menne2012a}.
Each station is identified with
a U.S.\ Cooperative Observer Network station identification code
which we will denote as Station ID.

The scatter plot in Fig.~\ref{fig:teletherm_map002},
along with all maps that follow,
demonstrate that the geographic coverage 
afforded by the stations is fairly uniform with some minor clustering
around populous areas. 

The temperature records in our data set are not, however, temporally uniform.
Stations have different lifespans---the 
oldest starting in 1853 
(Camden 3 W, South Carolina; Station ID: 381310)
and the youngest in 1998 
(Md Sci Ctr Baltimore, Maryland; Station ID: 185718).
Some stations have gaps in their
records, a complication which 
we will deal with as needed in our various analyses.
For example, 
Yellowstone Park in Mammoth, Wyoming
(Station ID: 489905)
has records for 1894--1903 and 1941--2012, missing a period of 37 years.
We will also ignore any datum for which a potential source of error 
is indicated.

In our analyses, we will use overall day number of the year for maximum
temperature starting at January 1.  
For the minimum temperature, we wrap the calendar and
consider days counting forwards from July 1 and running through to June 30 
in the following year, thereby roughly centering the low point to
better accommodate statistical treatment.
To present our results, we presume a 365 day year meaning 
an adjustment of a day will be needed for a leap year.

Finally, to create a reference tying the Teletherms to the solar cycle,
we standardize the Summer and Winter Solstices 
as falling on June 21 and December 21 (day numbers 172 and 355).

\section*{Analysis and Results}
\label{sec:teletherm.results}

\subsection*{Teletherms of Individual stations}
\label{subsec:teletherm.results-stations}

\begin{figure*}[tbp!]
      \centering
  \includegraphics[width=1\textwidth]{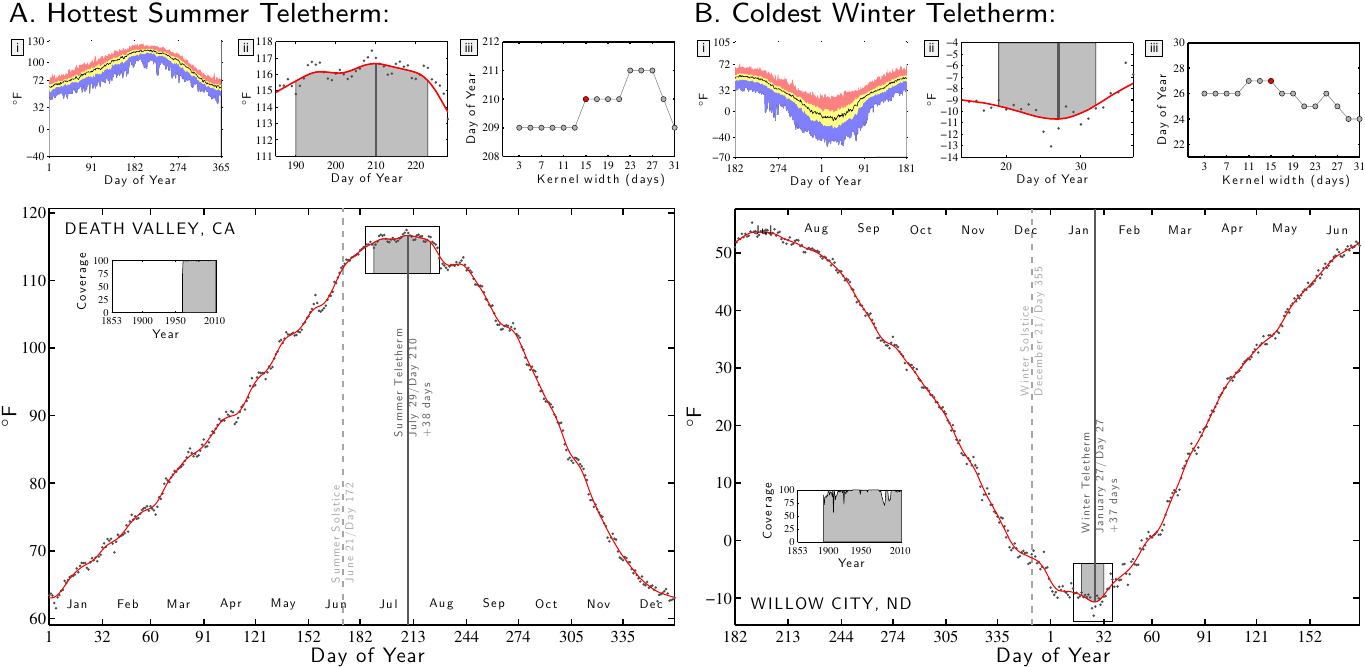}
  \caption{
    Plots establishing Teletherm date and Teletherm Periods for
    the examples of 
    \textbf{A:} the hottest summer Teletherm (Death Valley, California)
    and 
    \textbf{B:} the coldest winter Teletherm (Willow City, North Dakota).
    The main plots in \textbf{A} and \textbf{B} 
    show the average 
    daily maximum and minimum temperature (black dots)
    along with a smoothed curve formed using a Gaussian Kernel
    (solid red).
    For all minimum temperature analyses,
    we wrap the year
    from July 1 to June 30.
    The main plots' insets show the fraction of error-free recording
    for each year.
    Subplot \textbf{i:}
    Representation of the spectrum of maximum/minimum temperatures per day of the year.
    The black curve indicates the median,
    the blue area indicates lowest to first quartile, 
    yellow the
    inter-quartile range, 
    and red the fourth quartile.
    Subplot \textbf{ii:}
    Expansion of the inset around the Teletherm in the main plot.
    The dark gray vertical line indicates the Teletherm and the lighter gray
    region
    the Teletherm Period which we define
    as the days for which the smoothed maximum/minimum temperature curve
    is within 2\% of the Teletherm's temperature, 
    relative to the dynamic
    range of the smoothed curve over the entire 365 days.
    Subplot \textbf{iii:}
    Robustness diagnostic showing how the Teletherm date 
    varies as a function of Kernel width.
    We use 15 days, marked in red.
    See the main text for further details.
    See Fig.~\ref{fig:teletherm_tmax_station007_extremes002}
    for four more extreme Teletherm examples.
    We provide Teletherm plots for the maximum and minimum temperatures 
    for all 1218 stations in the Supporting
    Information (Files S22 and S23)
    and in the paper's online appendices at
    \protect\url{http://compstorylab.org/share/papers/dodds2015c}.
  }
  \label{fig:teletherm_tmax_station007_extremes001}
\end{figure*}

\begin{figure*}[tbp!]
      \centering
  \includegraphics[width=1\textwidth]{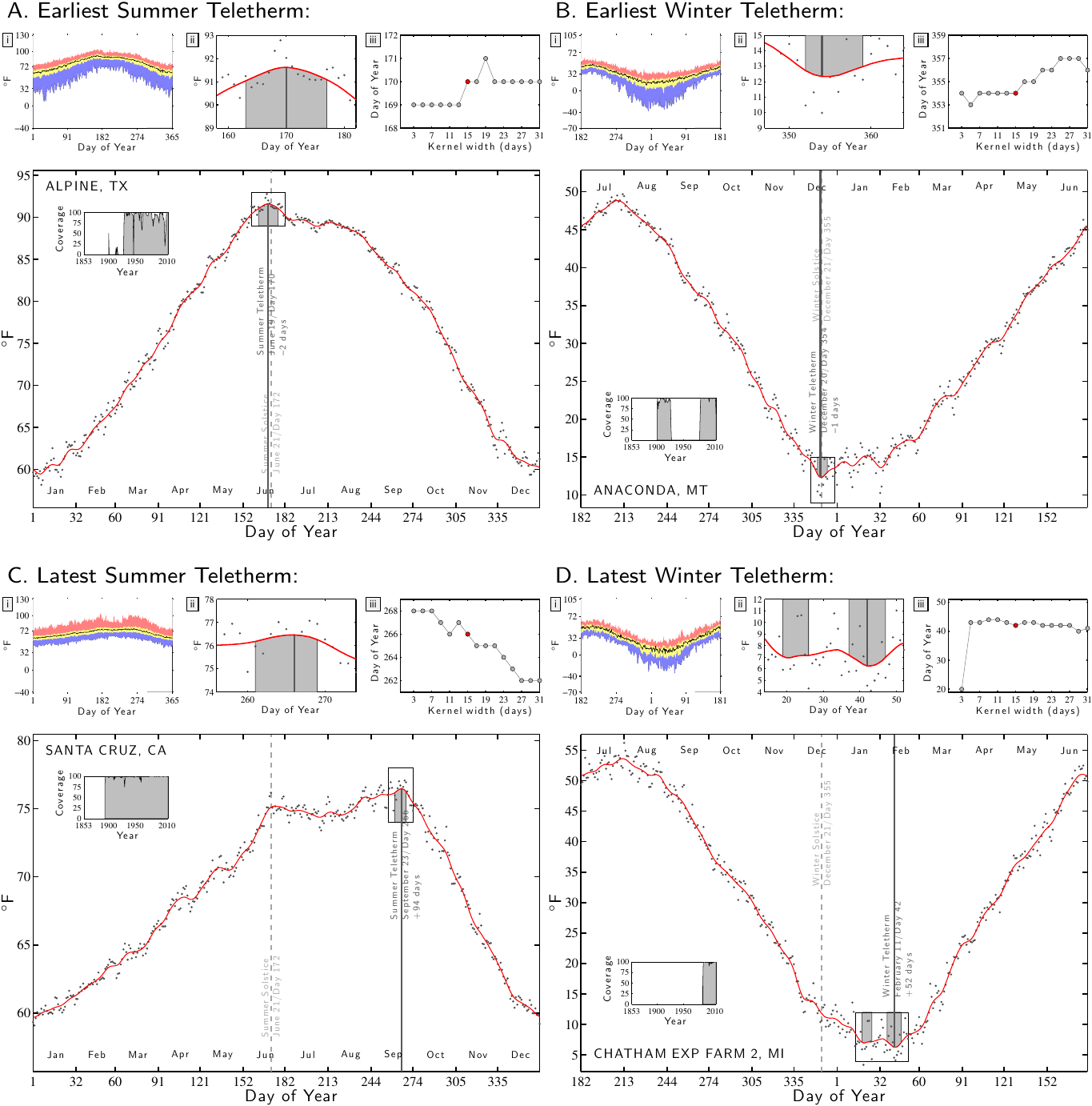}
  \caption{
    Teletherm plots for four extremes for the contiguous U.S.:
    the earliest Summer and Winter Teletherms 
    and
    the latest Summer and Winter Teletherms.
    See the caption of
    Fig.~\ref{fig:teletherm_tmax_station007_extremes001}
    for full details.
  }
  \label{fig:teletherm_tmax_station007_extremes002}
\end{figure*}

Our goal is to identify the Teletherms and Teletherm Periods 
and their respective dynamics in as straightforward a fashion as possible.
Because of the stochasticity of temperature, 
our analysis necessarily involves several steps.

We first compute the mean maximum and minimum temperature for each
day of the year at each station.
We average over all error-free data
points, acknowledging the variability of both length
and completeness of each station's temperature time series.
In the following section on Teletherm maps, 
we will only include averages for stations for which we have
data for at least 80\% of the dates within a given window.

To enable us to illustrate and explain our treatment in full,
we will use a selection of six extreme Teletherm locations 
in the contiguous U.S.
In Figs.~\ref{fig:teletherm_tmax_station007_extremes001}A--B
and~\ref{fig:teletherm_tmax_station007_extremes002}A--D,
we present diagnostic plots for the following specific Teletherms:
\begin{itemize}
\item 
  Fig.~\ref{fig:teletherm_tmax_station007_extremes001}A.
  Hottest Summer Teletherm: 
  Death Valley, California,
  (Station ID: 042319).
\item
  Fig.~\ref{fig:teletherm_tmax_station007_extremes001}B.
  Coldest Winter Teletherm:
  Willow City, North Dakota,
  (Station ID: 329445).
\item
  Fig.~\ref{fig:teletherm_tmax_station007_extremes002}A.
  Earliest Summer Teletherm: Alpine, Texas
  (Station ID: 410174).
\item
  Fig.~\ref{fig:teletherm_tmax_station007_extremes002}B.
  Earliest Winter Teletherm: Anaconda, Montana
  (Station ID: 240199).
\item
  Fig.~\ref{fig:teletherm_tmax_station007_extremes002}C.
  Latest Summer Teletherm: Santa Cruz, California.
  (Station ID: 047916).
\item
  Fig.~\ref{fig:teletherm_tmax_station007_extremes002}D.
  Latest Winter Teletherm: Chatham Exp Farm 2, Michigan
  (Station ID: 201486).
\end{itemize}

Each figure has the same format: 
a main plot showing average and smoothed
maximum or minimum temperature (explained below),
and three subplots across the top.
We will address the main plots first.

Taking the example of the Death Valley station,
in the main plot in 
Fig.~\ref{fig:teletherm_tmax_station007_extremes001}A,
the black dots represent the average maximum temperature for
each day of the year.
We smooth these points by convolving the average maximum
temperature time series with a Gaussian kernel of width 15 days, 
resulting in the red curve, and we elaborate on this
choice below.

After smoothing the data, we assign the day of the most extreme value of the 
resultant curve as the Teletherm for that station.
In all plots, we indicate Teletherms with a gray vertical line
and for reference, we locate the Summer or Winter Solstice
with a dashed gray vertical line.

The left inset in each main plot shows the fraction of days with
error-free data as a function of year.
In the case of Death Valley, we see the 
data set contains records from 1961 on,
and that these are fairly complete.
For Willow City in
Fig.~\ref{fig:teletherm_tmax_station007_extremes001}B,
the period of record begins before 1900 but shows
an imperfect collection rate; we generally see
that winter temperatures, especially minima,
are (unsurprisingly) more error prone.

Turning to the Teletherms themselves,
for Death Valley, we estimate that the
Summer Teletherm 
occurs on July 29 (day 210), 
a considerable 38 days after the Summer Solstice
(Fig.~\ref{fig:teletherm_tmax_station007_extremes001}A).
The coldest Winter Teletherm
occurs on January 27 in Willow City, North Dakota,
a similarly lengthy 37 days after the Winter Solstice
(Fig.~\ref{fig:teletherm_tmax_station007_extremes001}B).
While we define Teletherms as the date,
each one has of course an associated effective temperature
arising from our analysis.
For Death Valley,
this temperature is \tempfc{117}{47}
and for Willow City, 
we find \tempfc{-11}{-24}.
Death Valley also has the maximum temperature
recorded in the data set: \tempfc{129}{54}.

The earliest Teletherms occur in 
Alpine, Texas for the summer
(Fig.~\ref{fig:teletherm_tmax_station007_extremes002}A)
and 
Anaconda, Montana for the winter
(Fig.~\ref{fig:teletherm_tmax_station007_extremes002}B).
These Teletherms precede the adjacent Solstice by 
two days and one day respectively following a long linear change in temperature,
and both display an initially slow return afterwards.

The Teletherms occurring latest in the year 
have different stories.
For the summer, Santa Cruz's Teletherm is
experienced extremely late on September 23---essentially the Autumnal
Equinox---around  three months (94 days) after the Summer Solstice.
As Fig.~\ref{fig:teletherm_tmax_station007_extremes002}C
shows,
the average maximum temperature for Santa Cruz rises to a false
peak (a localized Teletherm) at the Summer Solstice, drops slightly and then climbs
again to the true Teletherm.
We find similar behavior for stations along the west coast
but not to any extent inland, a feature we examine 
further in the following section.

We estimate that the latest winter Teletherm takes place on February
11---a remarkable 52 days after the Winter Solstice and 9 days after Groundhog Day---at the 
Chatham Exp Farm 2 station in Michigan's Upper Peninsula
(Fig.~\ref{fig:teletherm_tmax_station007_extremes002}D).

We note that many of the smoothed average maximum and minimum
temperature curves we observe exhibit a
small periodic behavior as they climb and fall.
Not being a focus of our present work,
we suggest a more detailed analysis may uncover the source,
if any, of these apparent pulsings in the time series.

Continuing with our explanation of our analysis,
we move to the three diagnostic subplots marked \textbf{i}, 
\textbf{ii}, and \textbf{iii}
in each of 
Figs.~\ref{fig:teletherm_tmax_station007_extremes001}A--B and~\ref{fig:teletherm_tmax_station007_extremes002}A--D.
The first subplot \textbf{i}
summarizes the distribution
of maximum or minimum temperature for each station. 
The black curve gives the median for each day of the year,
the yellow region represents the inter-quartile range,
and the blue and red regions show the rest of the range.
For example, the top of the red region for a Summer Teletherm figure
indicates the hottest maximum temperatures, the bottom of the blue
the lowest maximum temperatures.
The stochasticity of the extreme temperatures measured
at the levels of day is readily apparent in these subplots.

The second subplot \textbf{ii} is an expanded and rescaled
match of the inset in the main plot around the Teletherm.
As for the main plot, the black dots show the average maximum or
minimum temperature for each day of the year, and the red curve
the smoothed version.
The gray shaded region shows the
full Teletherm Period for a station
which we describe below.

The third subplot \textbf{iii} shows how
the Teletherm varies as a function of 
the width of Gaussian kernel, providing
a measure of robustness.
To smooth the data, we used the Matlab command \textnormal{gausswin}
with Kernel width $W$ and standard deviation $\sigma = (W-1)/4$.
For the examples in
Figs.~\ref{fig:teletherm_tmax_station007_extremes001}A--B
and Figs.~\ref{fig:teletherm_tmax_station007_extremes002}A--D,
we see that the estimated date of the Teletherm varies 
relatively little---typically 2 to 4 days---for Kernel widths ranging from 7 to 31.

Our choice of a Gaussian kernel with a width of 15 is a
defensible, reasonable, and practical one, well within
what is a range of widths producing similar outputs and
interpretable as spanning a week to the side of each date.
We observe that very narrow kernels may however give quite different results 
as for the station Chatham Exp Farm 2 in
Fig.~\ref{fig:teletherm_tmax_station007_extremes002}D.
Such jumps may occur when two or more localized Teletherms are present
which we address in the next section.

\subsection*{Teletherm Periods for  Individual stations}
\label{subsec:teletherm.periods-stations}

In looking more closely at the behavior of average maximum
and minimum temperatures, we are obliged to augment
our definition of Teletherms beyond single days of the year.
Being able to assign one date to a location makes
for a simple story but we must acknowledge 
three aspects:
(1) We are working with a statistically speaking small number
of samples for each station;
(2) The choices we have made in our statistical analysis 
mean that the specific Teletherm date is subject to minor error;
and
(3) Fundamentally, some locations undergo on-average maximum
or minimum temperatures that hold over a range of dates.

We define the Teletherm Period for a location to be the range of dates,
possibly non-contiguous, for which the smoothed maximum/minimum
temperature curve lies within 2\% of the Teletherm's temperature
as measured with respect to the dynamic range of the smoothed curve.
We chose 2\% as a cutoff, asserting that the human-experienced
temperature would be roughly similar to that of the Teletherm.
An alternate approach would be to use an absolute difference
(e.g., within \tempf{1}); the results will not differ substantially.

Returning to Figs.~\ref{fig:teletherm_tmax_station007_extremes001}A--B
and Figs.~\ref{fig:teletherm_tmax_station007_extremes002}A--D,
we now identify the gray shaded region in
the inset around the Teletherm in the main plot
(reproduced in the subplot \textbf{ii})
as the Teletherm Period.
Across all stations, we see substantial variation in duration
and continuity of Teletherm Periods.
For Death Valley 
(Fig.~\ref{fig:teletherm_tmax_station007_extremes001}A)
the Teletherm Period lasts
an unpleasant 34 days with a smoothed maximum temperature
of at least \tempfc{115.6}{46.4}
(July 9th to August 11th, day numbers 190 to 223).
The Winter Teletherm Period for Anaconda, Montana
is comparatively brief running 
8 days with smoothed minimum temperatures below \tempfc{13.0}{-10.6}
(December 18th to 25th, numbers 352 to 359).

The station Chatham Exp Farm 2 in Michigan
(Fig.~\ref{fig:teletherm_tmax_station007_extremes001}D)
shows how our definition may lead to two or more Teletherm Periods 
surrounding minor cooling or warming periods.
In looking across all stations, we see that Winter Teletherms 
for stations in the Northeast may present a statistically sound early spring thaw,
and Burlington WSO AP, Vermont (Station ID: 431081) is another clear example
(see Supporting Information files S22 and S23 and
\url{http://compstorylab.org/share/papers/dodds2015c}).
Evidently, if we used a threshold of, say, 5\%, some separated
Teletherm Periods would coalesce, but we believe the threshold
should be suitably strict.

\begin{figure*}
  \centering
  \includegraphics[width=\textwidth]{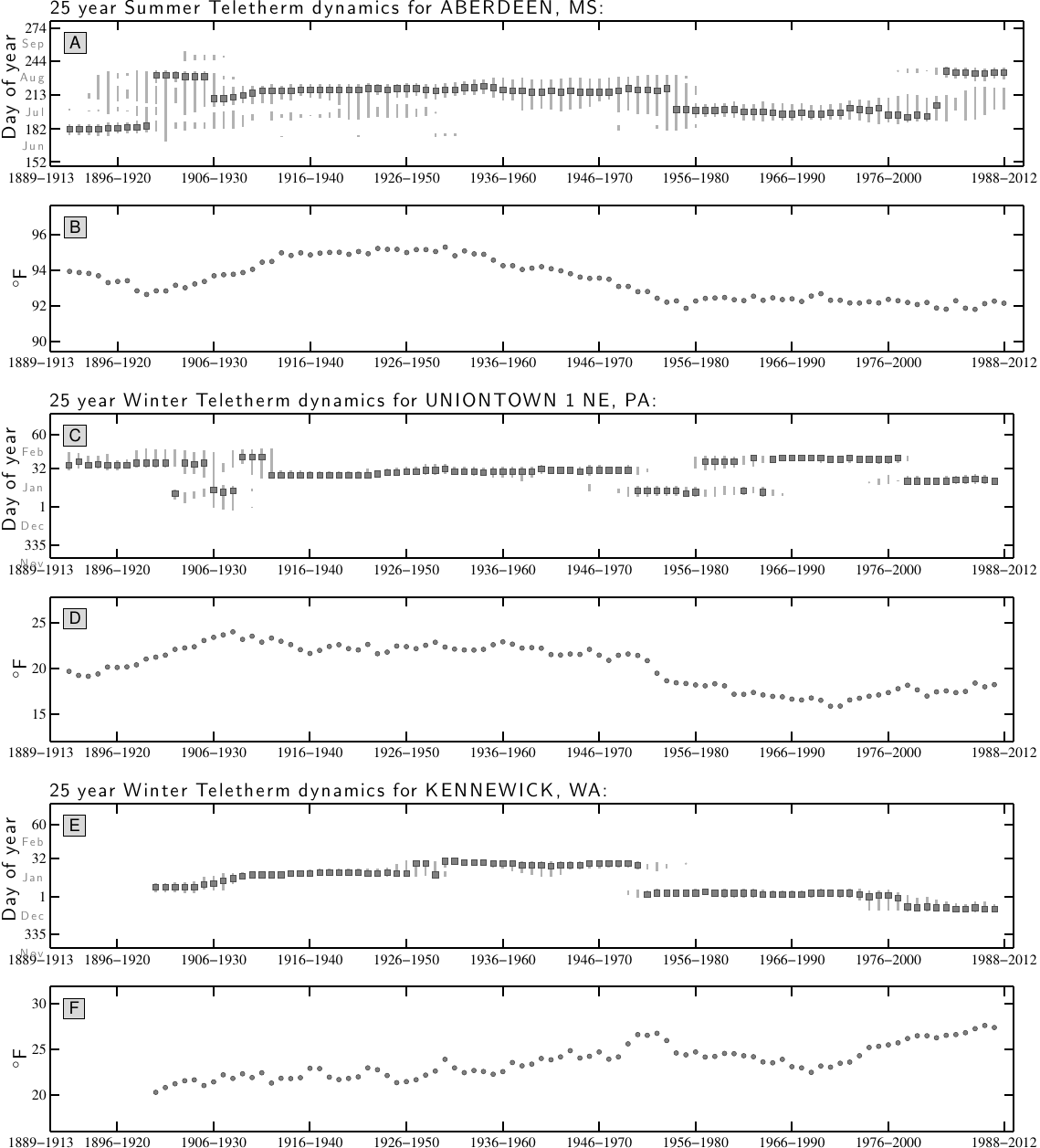}
  \caption{
    Dynamics of 25 year Teletherm dates, periods, extents, and temperatures
    for three example locations displaying abrupt switchings in time and
    gradual increases and decreases of temperature.
    We provide sets of these plots for 50, 25, and 10 year Teletherms
    for all 1218 stations in the Supporting Information (Files S28,
    S29, and S30).
    The same plots are also available at
    \protect\url{http://compstorylab.org/share/papers/dodds2015c/places.html}.
  }
  \label{fig:universal_teletherm_timelines002_025_both_examples}
\end{figure*}

For the whole data set, we observe considerable though locally coherent variation in
dynamics with temperatures rising and falling, and Teletherm periods
expanding, dividing, and coalescing, and Teletherm dates switching.
In Fig.~\ref{fig:universal_teletherm_timelines002_025_both_examples},
we show example behavior for 25 year Teletherms for 
Aberdeen, MS (Summer),
Uniontown, PA (Winter),
and
Kennewick, WA (Winter).
For Kennewick, we see the 25 year Winter Teletherm moves sharply
to an early date around the middle of the 20th century.
See Supporting Information Files S28, S29, and S30 for
the complete set of stations for 50, 25, and 10 year Teletherms.

\subsection*{Teletherm maps}
\label{subsec:teletherm.maps}

\begin{figure*}[tp!]
      \centering
      \includegraphics[width=0.90\textwidth]{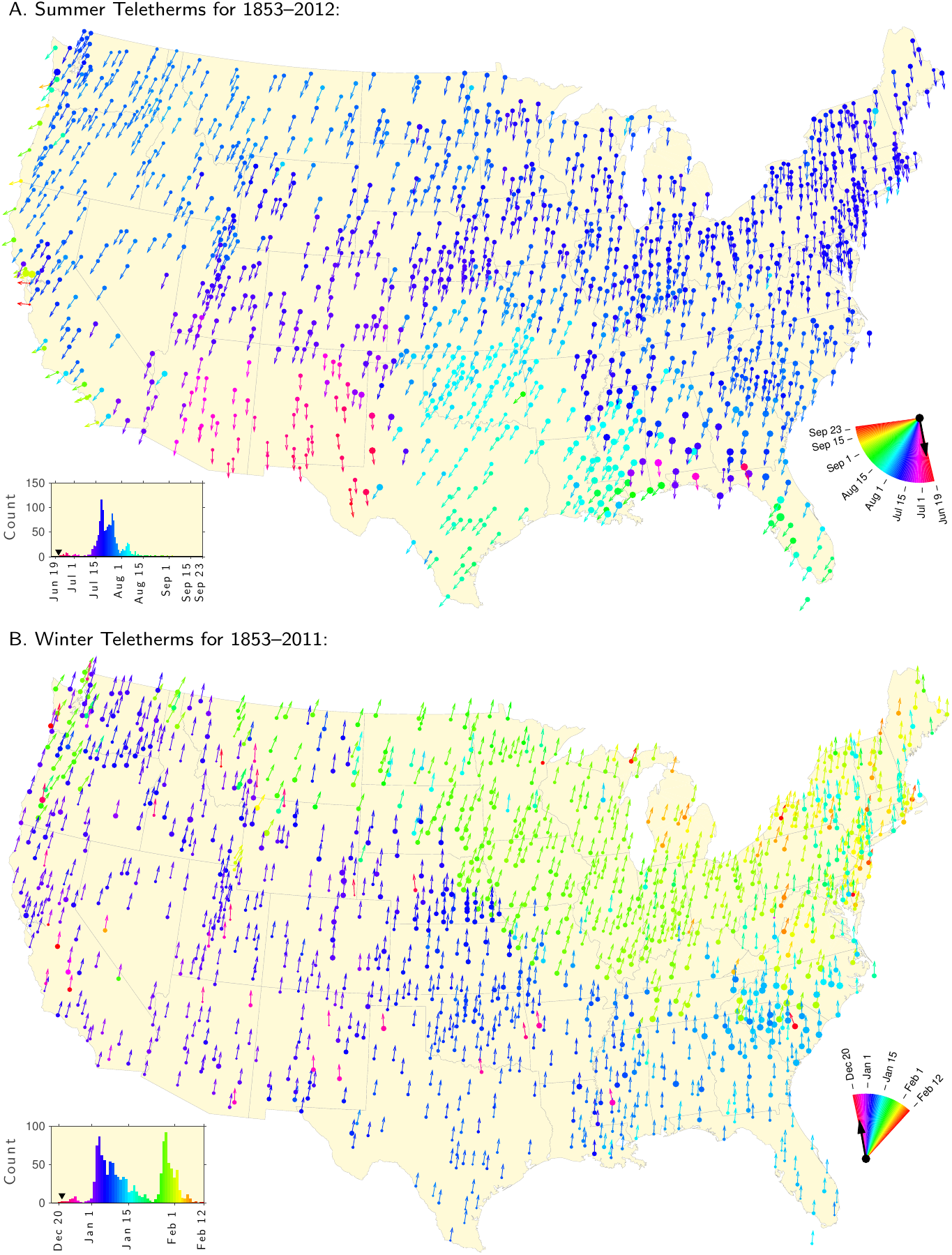}
      \caption{
        \textbf{A:} Summer Teletherms
        and 
        \textbf{B:} Winter Teletherms
        across the contiguous United States
        based on all data recorded from 1853 to 2012.
        Arrows point in the direction of the Teletherm's
        day of year mapped into angles traveling clockwise
        with December 31st aligned upwards.
        The sizes of the markers (discs) represent the duration of 
        a location's Teletherm Period.
        In the case of multiple Teletherm Periods, sizes
        correspond to the full extent.
        Colors map to Teletherm dates as indicated by the 
        partial color wheel in the bottom right corner of each map.
        The black arrows in the color wheels show the location of the Solstices.
        The histograms shows the distributions
        of the Summer and Winter Teletherm, using the same colors.
      }
      \label{fig:teletherm_range007_008}
\end{figure*}
 
We move now to exploring how the Teletherms vary across
the contiguous U.S. through maps. 
Once again drawing on the full data set,
we plot the 
Summer Teletherms  
in Fig.~\ref{fig:teletherm_range007_008}A
and
the 
Winter Teletherms
in Fig.~\ref{fig:teletherm_range007_008}B.
We present accompanying maps of the Teletherm temperatures
in Figs.~\ref{fig:teletherm_extremedays001}
and~\ref{fig:teletherm_extremedays002},
and a map showing the number of days
separating the two Teletherms at each station
in Fig.~\ref{fig:teletherm_diffs001}.

On all maps here and in the Supplementary Information, 
we indicate the Teletherm's day of year by an
arrow on a clock.
We orient the angle 0 radians upwards and assign days of the
standard year to multiples of $1/365 \times 2 \pi$
(December 31st then corresponds to angle 0).
To reinforce the visibility of variation, 
we color points and arrows per the color wheel
in the bottom right corner of all maps.
The black arrows in these color wheels mark 
the Summer and Winter Solstices as appropriate.

We visually supply information about the Teletherm Period by
linearly scaling the size of the marker for each location.
For stations with multiple Teletherm Periods, we use
what we call the Teletherm Extent---the number of days from the start
of the first Teletherm period to the end of the last one (inclusive).

We provide a histogram of the Teletherm days of the year
in the bottom left corner of each map, again using the same
color scheme.  The inverted black triangle identifies the 
relevant Solstice.

A number of observations stand out. 
For the Summer Teletherm, we see considerable but largely smooth
variation.
From Figs.~\ref{fig:teletherm_tmax_station007_extremes001}A
and \ref{fig:teletherm_tmax_station007_extremes001}C,
we had identified that the range of dates for the Summer Teletherm spans 96 days
(June 19 in Alpine, Texas to September 23 in Santa Cruz, California),
but we now see that the bulk of Teletherms fall between July 15
and August 1 (dark blue).
These second-half-of-July Teletherms
hold in the north of the contiguous U.S. and extend
down into California on the west and Georgia on the east.

The variant Summer Teletherms span several regions.
The earliest summer Teletherms occur in Arizona, New Mexico,
and the west of Texas (purple/red).  
In moving from west to east, we see a longitudinal discontinuity
in Texas with a switch to relatively late Summer Teletherms,
which remain apparent in Oklahoma, Arkansas, Louisiana,
Mississippi, and over to Florida.  
These August Teletherms form a noticeable minor peak in the histogram
(light blue).
The gulf coast shows some irregularity in the
Teletherm but more clearly exhibits the longest Teletherm Extents.

Stations along the west coast show how exposure
to the Pacific and incoming weather patterns make them break strongly with the nearby
inland Teletherm ``directions'', moving to generally later in the year
as per example of Santa Cruz we examined earlier (Fig.~\ref{fig:teletherm_tmax_station007_extremes002}C).
By contrast, stations along the east coast are consistently aligned
with their inland counterparts.

For the Winter Teletherm, we see a different overall pattern
with the contiguous U.S.\ dividing into two regions:
the west, midwest, and south with largely early January Winter
Teletherms (blue),
and the mid-north and northeast showing Teletherms in late January
and early February (green).
In the northeast's winter, the temperature continues to fall 
well beyond the shortest day of the year in the
northeast, typically 5 to 6 weeks.
We venture that a possible source of this clear regional separation 
might lie in the jet stream’s dynamics
across North America, with snowfall leading to increased albedo in the
northern section coupled with a continental-scale shadow of the
Rockies.
Beyond the scope of
the present analysis, future modeling would be needed to
properly test such an hypothesis.

\subsection*{Teletherm dynamics}
\label{subsec:teletherm.dynamics}

\begin{figure*}[tp!]
          \centering
      \includegraphics[width=0.90\textwidth]{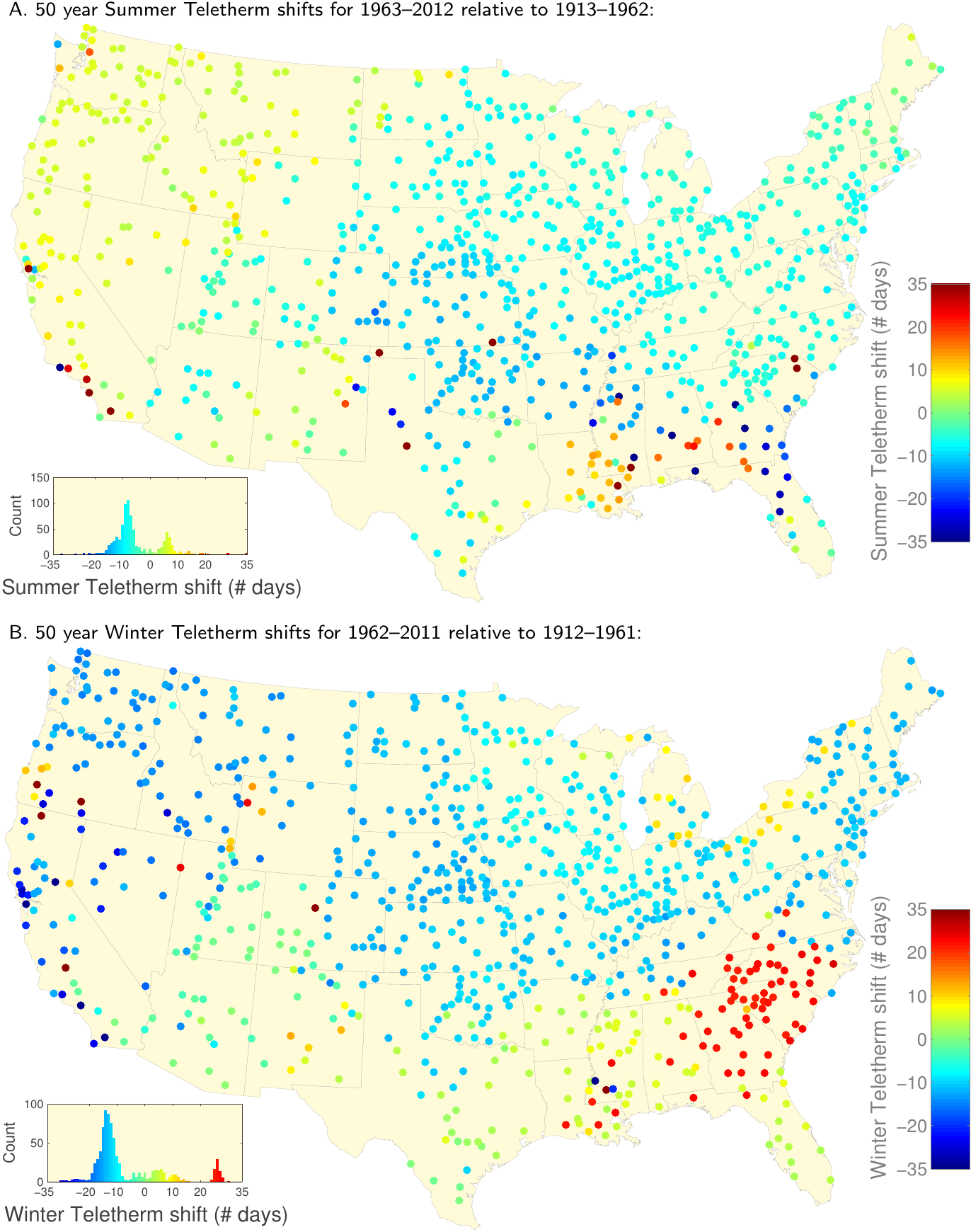}
      \caption{
        \textbf{A:}
        Summer Teletherm shifts comparing the 50 year period 1963--2012 relative to
        1913--1962.
        We show a total of 837 out of 1218 (68.7\%)
        which have $\ge$ 80\%
        error-free data in both 50 year spans.
        See Figs.~\ref{fig:teletherm.universal_teletherm_range_dynamic_tmax001_050_1913_to_1962}
        and~\ref{fig:teletherm.universal_teletherm_range_dynamic_tmax001_050_1963_to_2012}
        for maps of the Summer Teletherms for each period.
        \textbf{B:}
        Winter Teletherm shifts comparing 1962/1963--2011/2012
        relative to 1912/1913--1961/1962.
        A total of 835 out of 1218, 68.6\%, 
        stations have $\ge$ 80\%
        error-free data. 
        See
        Figs.~\ref{fig:teletherm.universal_teletherm_range_dynamic_wrapped_tmin001_050_1912_to_1961}
        and
        \ref{fig:teletherm.universal_teletherm_range_dynamic_wrapped_tmin001_050_1962_to_2011}
        for maps of the Winter Teletherms for each period.
      }
      \label{fig:teletherm_changes001_max_min}
\end{figure*}

In order to discern Teletherm dynamics and their
potential value in quantifying and studying climate change, 
we carry out the same smoothing we have performed
for the full time range for sliding windows of 
50 years in duration.
We also now make our data requirements more stringent
and estimate the Teletherm and
the Teletherm Period(s) for only those time ranges for which
we have 80\% of all temperatures recorded.

Here, we present and discuss shifts in Summer and Winter Teletherm dates for
two example consecutive 50 year periods: 1913--1962 and 1963--2012
(the six month offset leads to references to one year earlier for the Winter Teletherm).
We include all related Teletherm maps in the Supporting Information 
in the form of PDF flipbooks (files S24, S25, S26, and S27).
We also provide interactive visualizations of 
Teletherm dynamics at
\url{http://compstorylab.org/share/papers/dodds2015c}
and
\url{http://panometer.org/instruments/teletherms}.

In Fig.~\ref{fig:teletherm_changes001_max_min}A, we show how
the Summer Teletherm has moved between these two
half century time periods.
We compute the ``Teletherm shift'' in days
(see Figs.~\ref{fig:teletherm.universal_teletherm_range_dynamic_tmax001_050_1913_to_1962}
and~\ref{fig:teletherm.universal_teletherm_range_dynamic_tmax001_050_1963_to_2012}
for plots of the respective Summer Teletherms)
and use a color map to present the results.
The lower left histogram in Fig.~\ref{fig:teletherm_changes001_max_min}A
represents the distribution of shifts.
Now, if change was random, we would expect to see a normal 
distribution centered around a shift of 0.
We instead find two peaks separated from zero shift,
meaning very few locations experienced no change.
The larger peak (blue) means the Summer Teletherm has moved
to earlier in the year, connecting more strongly with the Solstice, 
and corresponds generally to
the northeast extending across and down into the midwest and south.
Stations in the west are reflected in the histogram's 
smaller peak (green/yellow) indicating the Summer Teletherm has
moved to later in the year for that area's stations.

We show shifts in the Winter Teletherm in
Fig.~\ref{fig:teletherm_changes001_max_min}B
(see
Figs.~\ref{fig:teletherm.universal_teletherm_range_dynamic_wrapped_tmin001_050_1912_to_1961}
and~\ref{fig:teletherm.universal_teletherm_range_dynamic_wrapped_tmin001_050_1962_to_2011}
for the Teletherms themselves).
We again find a texture different to that of 
the Summer Teletherm.
The dominant change is that the Winter Teletherm has advanced to earlier dates
in the year across the northern half of the contiguous U.S.,
and down along the west coast (blue).
As the histogram shows, the spread of forward shifts peaks in the range 10 to 20 days.
Going in the other direction, we see that 
the Winter Teletherms for the states of Georgia and the two Carolinas
have experienced a delay of around 25 days (red).
For the rest of the contiguous U.S., from New Mexico across
to Florida, the Winter Teletherm has remained fairly constant.
Comparing the histograms for the Winter Teletherm dates
in Figs.~\ref{fig:teletherm.universal_teletherm_range_dynamic_wrapped_tmin001_050_1912_to_1961}
and~\ref{fig:teletherm.universal_teletherm_range_dynamic_wrapped_tmin001_050_1962_to_2011},
we see that three distinct peaks have merged into one grouping over
time.  In sum, the Winter Teletherm has become more homogeneous in the
eastern half of the contiguous U.S., with both early and late Winter
Teletherms moving into the first half of January, while largely moving to earlier
dates in the west.

In Figs.~\ref{fig:universal_teletherm_summer_changes_temp_extents}A--B
and~\ref{fig:universal_teletherm_winter_changes_temp_extents}A--B,
we present the shifts in Teletherm Temperatures and Extents
for the same pair of 50 year periods.
The changes in temperature are milder for the Summer Teletherm 
($\pm$ \tempf{5},
Fig.~\ref{fig:universal_teletherm_summer_changes_temp_extents}A) 
than for the 
Winter ($\pm$ \tempf{10},
Fig.~\ref{fig:universal_teletherm_winter_changes_temp_extents}A).
The Extents however have maximally increased or decreased by 30 to 40
days,
with the Summer Teletherm seeing the most flux.
(Figs.~\ref{fig:universal_teletherm_summer_changes_temp_extents}B
and~\ref{fig:universal_teletherm_winter_changes_temp_extents}B).
Some of the other trends we see are that 
(1) the Summer Teletherm Temperature has dropped in the middle of the
contiguous United States while remaining neutral or increasing
elsewhere;
(2) Summer Teletherm Extents have increased most strongly throughout
the south;
(3) The Winter Teletherm Temperature has lowered in the South East
and increased in the central and western areas of the north;
and
(4) Winter Teletherm Extents have decreased in the south and
increased in areas around the Great Lakes.

Finally, we observe that the transitions in Teletherm features
between these two adjacent 50 year periods is not linear,
and that window length matters~\cite{trenberth2015a}.
To show this, we break the same century (1913--2012) into four 25 year periods.
First, we see a strengthened version of the 
same general overall changes to the Teletherm dates 
as for the 50 year analysis 
in comparing the last 25 years to the first 25 years (1988--2012
relative to 1913--1937)
(Fig.~\ref{fig:teletherm_changes002_max_min_1}).
The three transitions between the four 25 year periods show
accelerations, stasis, and reversals 
(see Figs.~\ref{fig:teletherm_changes002_max_min_2},
\ref{fig:teletherm_changes002_max_min_3},
and \ref{fig:teletherm_changes002_max_min_4}).
For example, in Louisiana and Mississippi,
the Summer Teletherm has shifted to later dates 
but through an advance, retreat, advance movement
(Figs.~\ref{fig:teletherm_changes002_max_min_2}A,
\ref{fig:teletherm_changes002_max_min_3}A,
and~\ref{fig:teletherm_changes002_max_min_4}A).
Much of the shift toward an earlier Winter Teletherm
across the north occurred in the 50 year period 1937--1986
(Fig.~\ref{fig:teletherm_changes002_max_min_3}B),
and the southeast first saw the Winter Teletherm advance
and then start to fall back to later dates
(Figs.~\ref{fig:teletherm_changes002_max_min_2}B,
\ref{fig:teletherm_changes002_max_min_3}B,
and~\ref{fig:teletherm_changes002_max_min_4}B).

We show the corresponding maps for shifts in Teletherm temperatures
and extents
in
Figs.~\ref{fig:teletherm_changes_temperature002_max_min_1},
\ref{fig:teletherm_changes_temperature002_max_min_2},
\ref{fig:teletherm_changes_temperature002_max_min_3},
\ref{fig:teletherm_changes_temperature002_max_min_4},
\ref{fig:teletherm_changes_extents002_max_min_1},
\ref{fig:teletherm_changes_extents002_max_min_2},
\ref{fig:teletherm_changes_extents002_max_min_3},
and
\ref{fig:teletherm_changes_extents002_max_min_4}.
The transition from the first 25 years to the last 25 years of
1913--2012
sees an average drop in the 25 year Summer Teletherm temperature
(mainly due the interior states)
but an increase in the Winter Teletherm temperature
(concentrated more along the north and down into Utah, Colorado,
and Arizona).
Of many notable details, we see a dropping of the Winter Teletherm's
temperature, in the eastern half
of the contiguous U.S. between 1937--1961 and 1962--1986,
followed by a reverse swing upwards over the next 25 years
(Figs.~\ref{fig:teletherm_changes_temperature002_max_min_3}B
and~\ref{fig:teletherm_changes_temperature002_max_min_4}B).

Interpreting the dynamics of the Teletherms is not an easy task
and we limit our assertions in this initial work.
We might suspect the jet stream may have played a part in the transition
of the Winter Teletherm in the southeast.  
Even without a clear understanding, we can see that
impact of these changes is potentially dramatic.
The movement of the Winter Teletherm for example alters
the local advent of spring, a strong driver of ecological systems.

\section*{Comparison to models}
\label{sec:teletherm.models}

\begin{figure}[tpb!]
          \centering
  \includegraphics[width=0.495\textwidth]{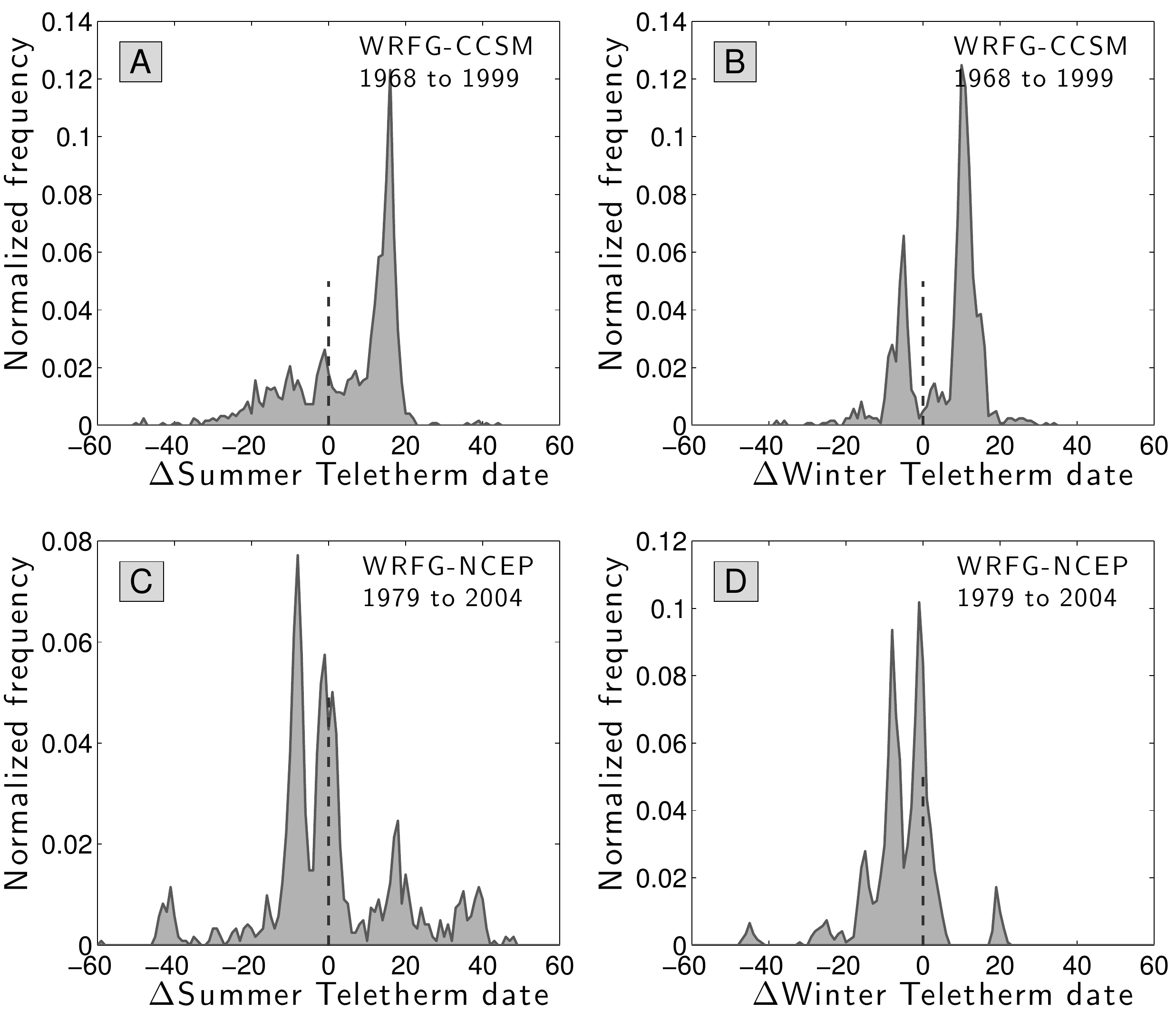}
  \caption{
    Distributions of errors in days for the Summer and Winter
    Teletherms at all stations
    when comparing measured temperatures
    to 
    the Community Climate Systems Model (CCSM) (version 3)~\cite{collins2006a}
    and 
    the  National Centers for Environmental Prediction Climate Forecast
    System Model (NCEP)~\cite{mearns2009a,mearns2014a}.
    The differences are to be interpreted as how many
    days the models are ``off'' from the real data.
    A positive $\Delta$  means the model's Teletherm
    occurs later in the year than the measured one.
  }
  \label{fig:modelerrordist}
\end{figure}

We end our analysis with a comparison of estimated Teletherm data
to output from two Regional Climate Models (RCMs) from the North
American Regional Climate Change Assessment Program (NARCCAP)~\cite{lempert2015a}.
Specifically, we analyze output of the WRF model nested within
both the Community Climate Systems Model (CCSM) (version
3)~\cite{barkemeyer2015a}
and
the National Centers for Environmental Prediction Climate Forecast
System Model (NCEP)~\cite{mearns2009a,mearns2012a,mearns2014a}. 
Details on the data can be found at
\href{https://www.earthsystemgrid.org/project/narccap.html}{https://www.earthsystemgrid.org/project/narccap.html}.

We use daily temperature extremes from both model systems at all 1218
station locations to compute Summer and Winter Teletherms for the time
periods covered by the models: 1968 to 1999 (CCSM) and 1979 to 2004
(NCEP) (see~\cite{greasby2012a} for related work on climate models).

Using our historical data set,
we also determined the Teletherms for these same time periods.
We then found the difference between the models' Teletherms 
and the measured Teletherms at each location,
and we show the histogram of these differences 
in Fig.~\ref{fig:modelerrordist}.
In these plots, a positive difference means a model's Teletherm
occurs later in the year than the Teletherm we estimated
based on real data.

For both models and for both Winter and Summer Teletherms,
we see evidence of characteristic, irregular kinds of errors.
For the CCSM, the single peak 
in Fig.~\ref{fig:modelerrordist}A shows 
that the model produces Summer Teletherms 
that occur 10 to 20 days later in the year than those observed.
For the Winter Teletherm, two peaks reflect
regional systematic errors.
The NCEP model fares somewhat better with a peak
around a difference of 0 for both Teletherms,
though a second peak of similar size indicates
a prediction of earlier Teletherms for a commensurate
swathe of stations.

We find the average absolute error in estimating
the Summer and Winter Teletherm are 12.88 and 10.05 days for the CCSM,
and 12.24 and 7.57 days for the NCEP model.
Spearman correlations are mixed with a best
value of 0.85 for the CCSM's Winter Teletherm ($p$-value effectively 0)
and a worst case of 0.059 for CCSM's Summer Teletherm ($p$-value 0.039).
At the level of stations, the worst errors for both models are for the Summer Teletherm 
with spans 78 and 59 days too early and 44 and 48 too late 
for the CCSM and the NCEP model respectively.

\begin{figure}[tbp!]
          \centering
  \includegraphics[width=0.495\textwidth]{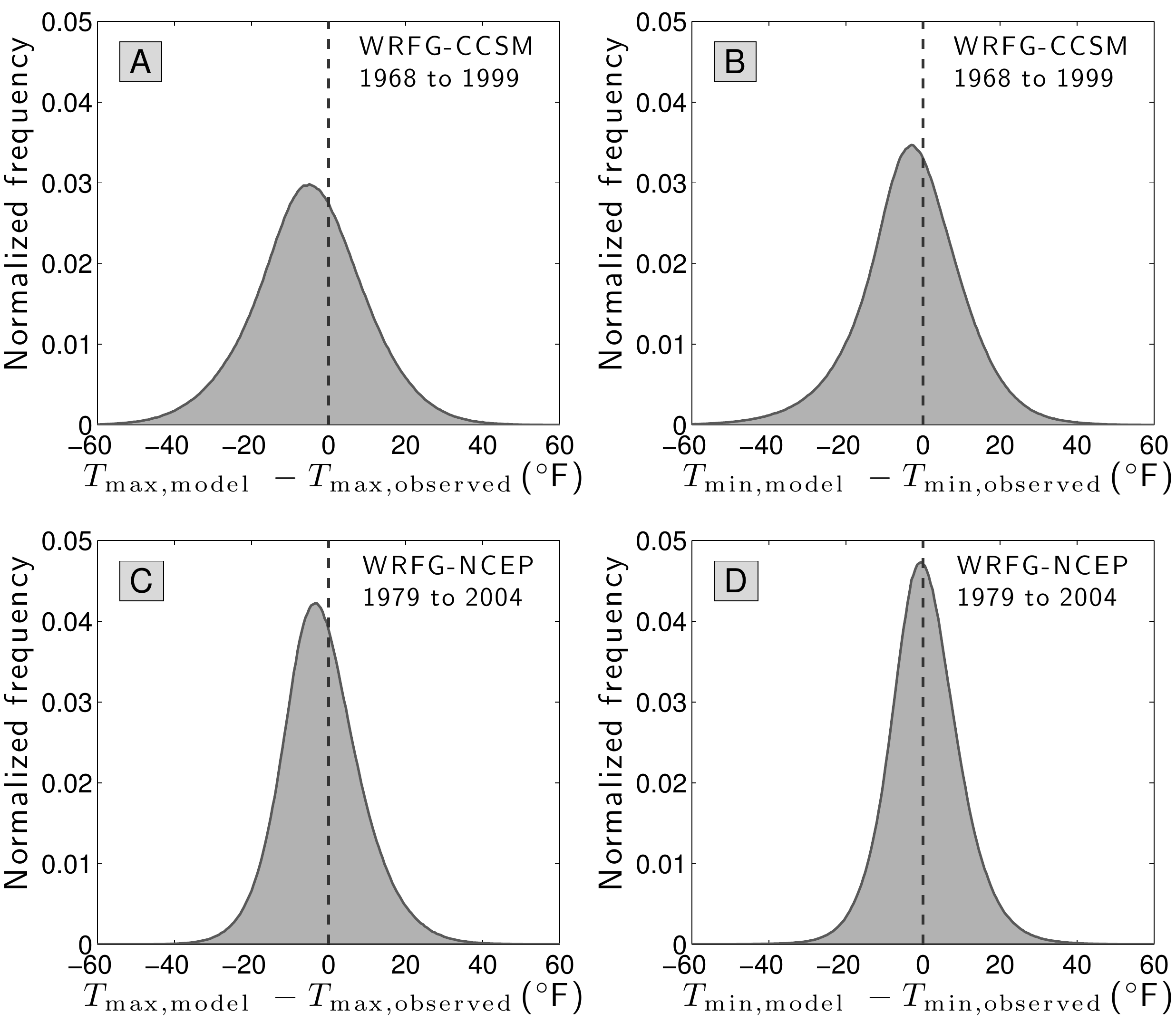}
  \caption{
    Comparison between predicted daily maximum and minimum
    temperatures generated by two climate models (CCSM and NCEP) 
    relative to
    real measurements for all stations.
  }
  \label{fig:modelerrordist_daily}
\end{figure}

In Fig.~\ref{fig:modelerrordist_daily}, we step back
from Teletherms, and 
plot the distribution of errors at the day level
between the output of both models and measured maximum and minimum
temperatures.
This is an exacting test: how does a model fair
with predicting the maximum temperature, say, in Death Valley
on March 3, 1982, along with all other stations and all other dates
over several decades?
With approximately 10,000 points per panel,
we see a much smoother distribution and the
form is now Gaussian-like.

We find that the Spearman correlations between the models' outputs
and measured daily temperature extremes are good,
ranging from 0.796 (CCSM, daily minimum temperature) to 0.875 (NCEP,
daily maximum temperature)
($p$-values effectively 0).
The NCEP model is on average more accurate with an average difference for
the daily minimum of \tempf{0.53}.  The average absolute error
varies from 7.32 (NCEP, daily minimum temperature) to 12.1 
(CCSM, daily maximum temperature).
The potential for wild inaccuracies remain with CCSM's worst prediction
being \tempf{95} below the real measurement of a minimum temperature.

In testing these climate models for Teletherm timing and daily
temperature extremes, we are certainly asking for more than they have
been intended to deliver.
Indeed, if these models were integrating in Numerical Weather
Prediction mode, with initial values updated through data
assimilation, the errors would be much smaller.
Nevertheless, understanding the successes and limitations of any model, 
whether aimed for or not, should be of benefit to future refinements~\cite{greasby2011a}.

\section*{Concluding remarks}
\label{sec:teletherm.conclusion}

We were initially motivated by the simple question of
when should we expect the on-average warmest and coldest day of the
year to occur at a given location.
In the northeast of the U.S.\ for example, the Winter Solstice
passes and as the days lengthen, the cold deepens
and people begin to wonder when will the winter end.
Traditionally, prognosticators have used diverse methods to divine
the length of winter such as, famously, how certain species of rodents react 
to their umbra. And in general, people look for signs of all the seasons
arriving such as the emergence of daffodils in spring 
or the first leaves turning to their autumnal colors.
We realized however that a data-driven, less poetic path could be assayed.

While the analysis promised to be initially straightforward
(as is often believed to be the case),
we soon found that we had to move beyond a single day version of the
Teletherm to a Teletherm Period.
Overall, we believe we have shown the spatiotemporal variability of the Teletherms 
and the surrounding Teletherm Periods to be considerable, informative,
and of general interest.
Importantly, we have seen that the variations in Teletherm characteristics are not 
a reflection of random noise but rather linear movements, periods of
stasis, and switching reminiscent of bifurcations in dynamical systems.
Teletherms seem therefore to present a real facet of climate change,
whatever the origin.

A number of future directions are possible.
Where data is available, our analysis could readily be carried out
for other regions around the world.
Beyond local interest, such efforts could 
lead to an effort to patch together a global picture of the 
Teletherms.
Online displays of Teletherms could also eventually include
the ability to adjust time frames for the analysis
and to show the likelihood that the warmest or coldest day has
occurred as a function of day of the year.
A global map would also afford more opportunities to
test models and hypotheses regarding climate dynamics.
For example, does the temporal behavior of Teletherms correlate in an
way to changes or stationarity of average annual temperature?
The stochastic nature of temperature could also be of
value in our collective general education  
regarding prediction for noisy systems.

We close by venturing that a region's Teletherm may also be acknowledged
annually (using, say, the most recent 50 years), potentially with a
set of associated food-based rituals or celebrations.

\acknowledgments
The authors appreciate helpful discussions with 
Istvan Szunyogh, Linda Mearns, John Kaehny, Bill Gottesman, Bruce Shaw, and Andrew Gelman.
The authors thank the North American Regional Climate Change
Assessment Program (NARCCAP) for providing the model data used in this
paper.
NARCCAP is funded by the National Science Foundation (NSF), the
U.S. Department of Energy (DoE), the National Oceanic and Atmospheric
Administration (NOAA), and the U.S. Environmental Protection Agency
Office of Research and Development (EPA).
LM, AJR, and CMD were in part supported
by the Mathematics and Climate Research Network (MCRN),
NSF Award \# DMS-0940271.
PSD was supported by NSF CAREER Award \# 0846668.

\clearpage

\newwrite\tempfile
\immediate\openout\tempfile=startsupp.txt
\immediate\write\tempfile{\thepage}
\immediate\closeout\tempfile

\setcounter{page}{1}
\renewcommand{\thepage}{S\arabic{page}}
\renewcommand{\thefigure}{S\arabic{figure}}
\renewcommand{\thetable}{S\arabic{table}}
\setcounter{figure}{0}
\setcounter{table}{0}

\noindent
\textbf{Supporting Information} for 

\bigskip
\noindent
\textbf{\protect{}Tracking climate change through the spatiotemporal dynamics 
of the Teletherms, the statistically hottest and coldest days of the year
}

\bigskip
\noindent

\bigskip
\noindent
Peter Sheridan Dodds,\\
Lewis Mitchell,\\ 
Andrew J.\ Reagan,\\ 
and Christopher M.\ Danforth.

\bigskip
\noindent
See also the paper's online appendices at:\\ 
\url{http://compstorylab.org/share/papers/dodds2015c}.

\begin{figure*}[h!]
      \centering
      \includegraphics[width=0.85\textwidth]{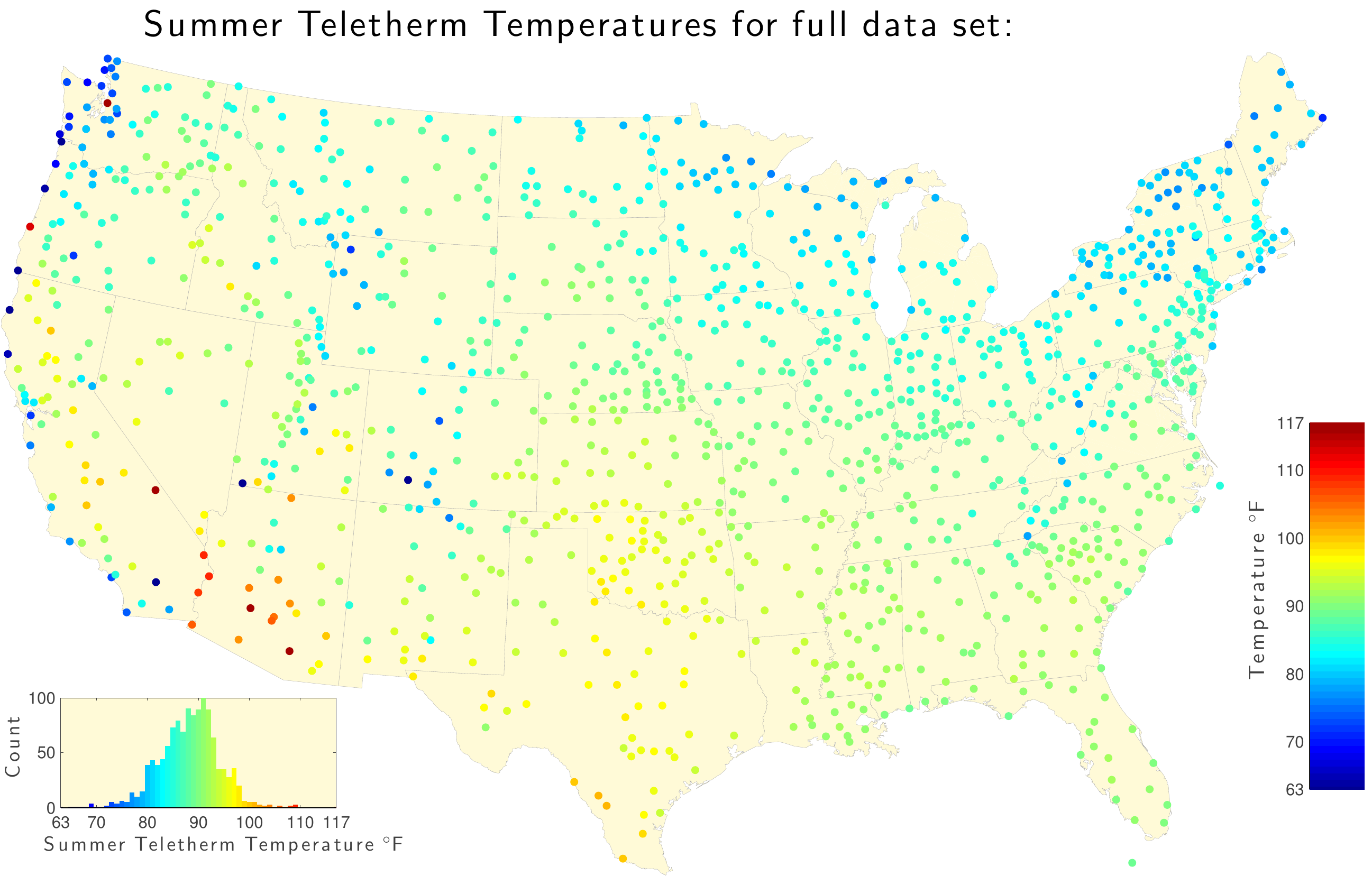}
      \caption{
        Summer Teletherm temperatures for the full data set
        (1853--2012).
        Teletherm temperatures are determined by smoothing the average
        daily maximum and minimum temperatures; see main text for details.
      }
      \label{fig:teletherm_extremedays001}
\end{figure*}

\begin{figure*}[h!]
      \centering
      \includegraphics[width=0.85\textwidth]{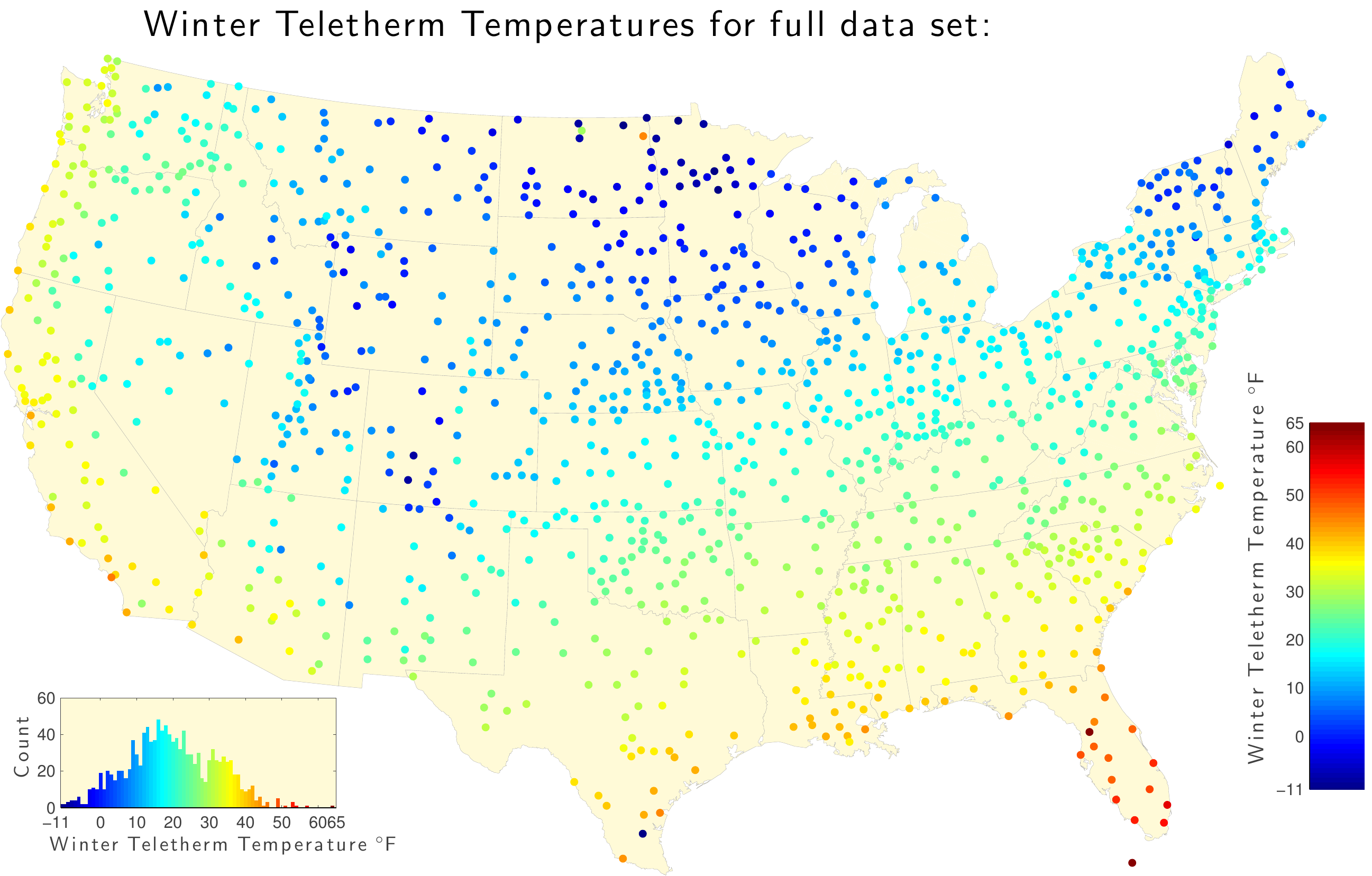}
      \caption{
        Winter Teletherm temperatures for the full data set (1853--2012).
        Teletherm temperatures are determined by smoothing the average
        daily maximum and minimum temperatures; see main text for details.
      }
      \label{fig:teletherm_extremedays002}
\end{figure*}

\begin{figure*}[h!]
      \centering
      \vspace{75pt}
      \includegraphics[width=0.85\textwidth]{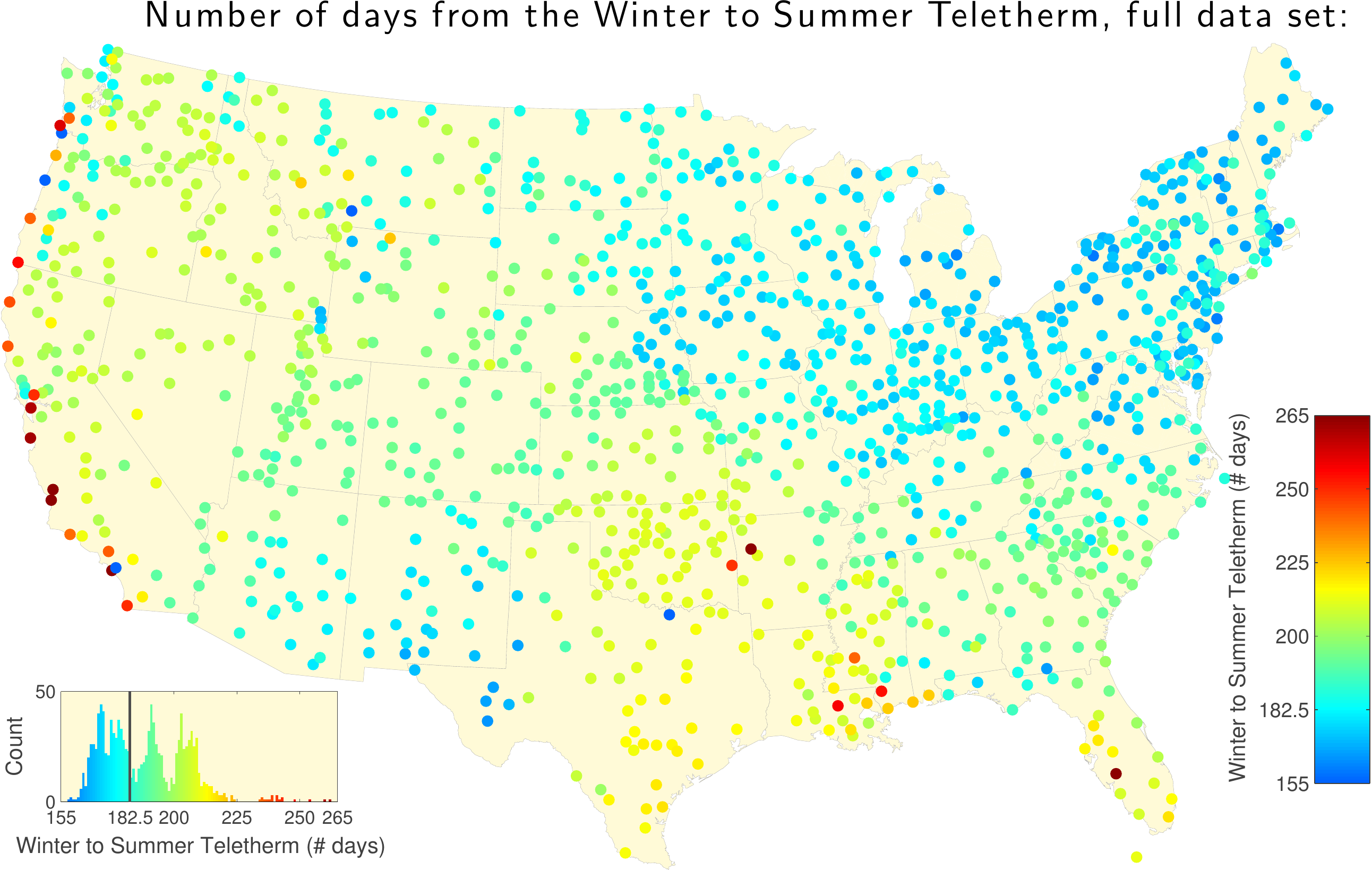}
      \caption{
        Number of days from the Winter to the Summer Teletherm.
        The vertical gray line in the histogram indicates half 
        of a standard 365 day year.
        The variation is substantial with the northeast showing
        as short a span as just over 5 months and the west coast as
        much as 9 months.
      }
      \label{fig:teletherm_diffs001}
\end{figure*}

\begin{figure*}[h!]
  \centering
            \includegraphics[width=0.85\textwidth]{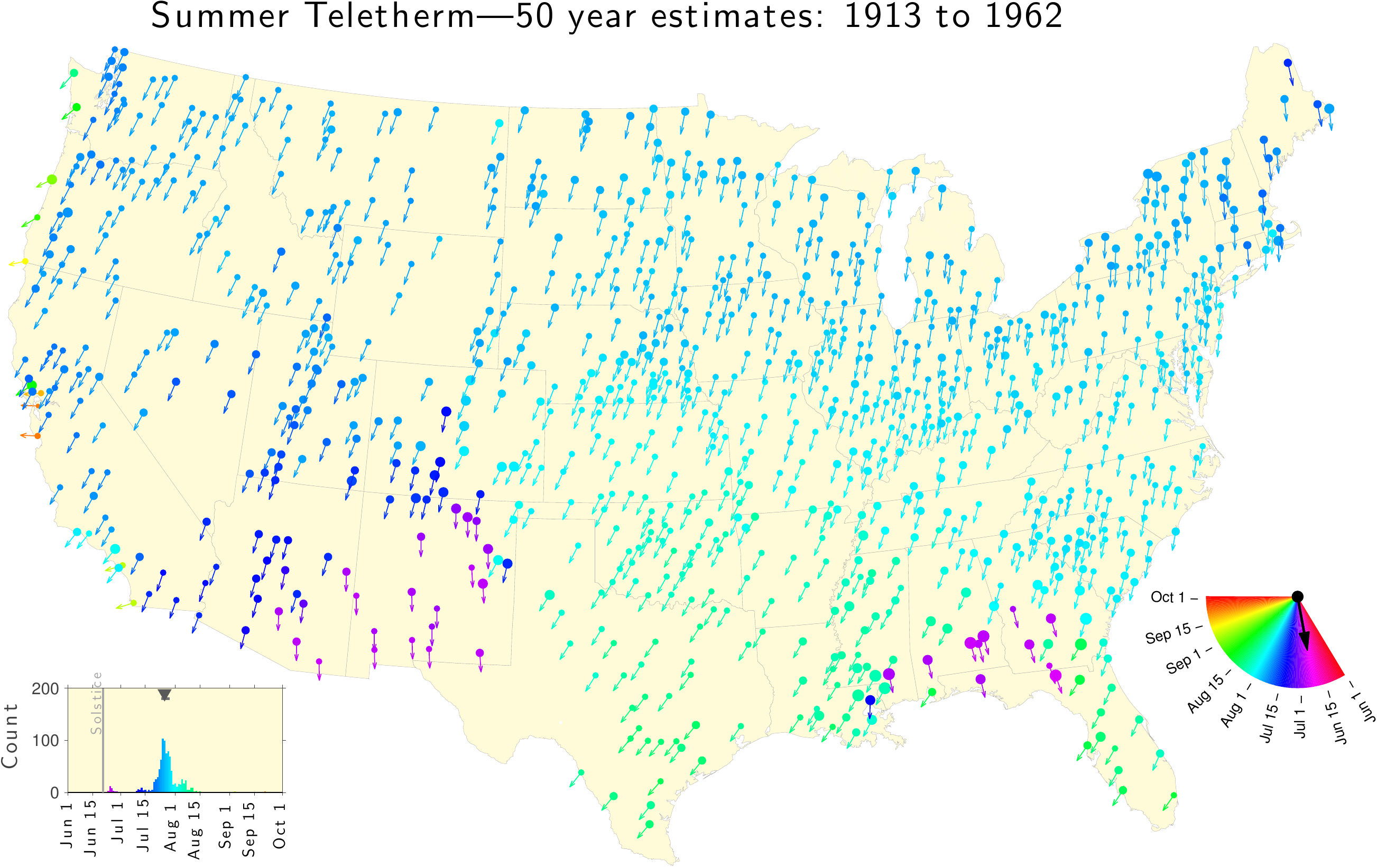}
  \caption{
    Map of the Summer Teletherms and Teletherm Extents estimated for the 50 year
    range 1913--1962,
    to be compared with the equivalent map for 1963--2012 in
    Fig.~\ref{fig:teletherm.universal_teletherm_range_dynamic_tmax001_050_1963_to_2012}.
    Fig.~\ref{fig:teletherm_changes001_max_min}A in the main text maps the
    changes
    in Summer Teletherms between these two periods.
    Relatively few Summer Teletherms have remained stable with
    the majority shifting to an earlier date.
    In the bottom left histograms, the gray horizontal line
    shows the interquartile range and the inverted triangle the median.
  }
  \label{fig:teletherm.universal_teletherm_range_dynamic_tmax001_050_1913_to_1962}
\end{figure*}

\begin{figure*}[h!]
  \centering
            \includegraphics[width=0.85\textwidth]{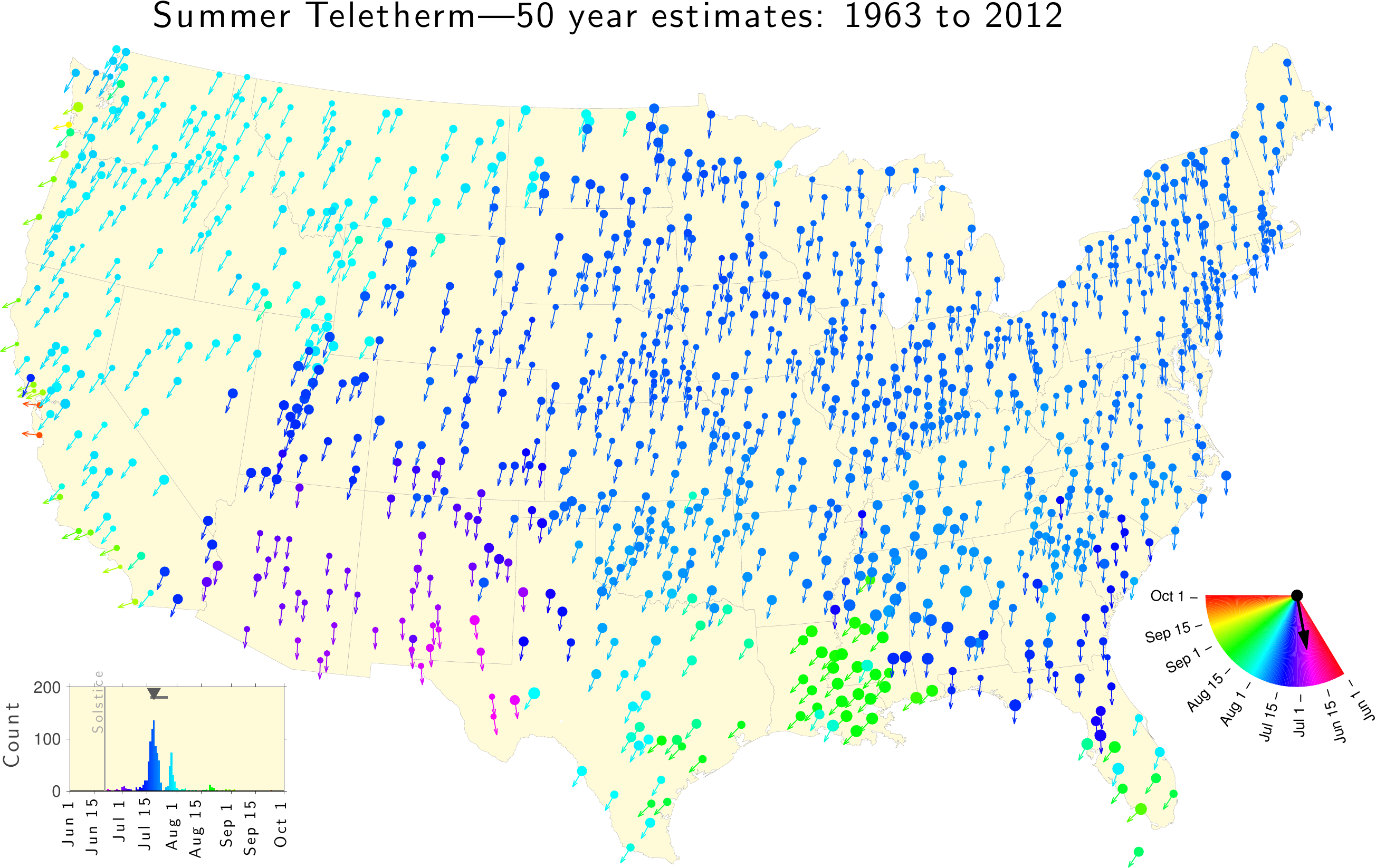}
  \caption{
    Map of the Summer Teletherms and Teletherm Extents estimated for the year
    ranges 1963--2012,
    to be compared with the preceding map in
    Fig.~\ref{fig:teletherm.universal_teletherm_range_dynamic_tmax001_050_1913_to_1962}.
    Fig.~\ref{fig:teletherm_changes001_max_min}A in the main text maps the
    changes in the Summer Teletherm between these two periods.
  }
  \label{fig:teletherm.universal_teletherm_range_dynamic_tmax001_050_1963_to_2012}
\end{figure*}

\begin{figure*}[h!]
  \centering
            \includegraphics[width=0.85\textwidth]{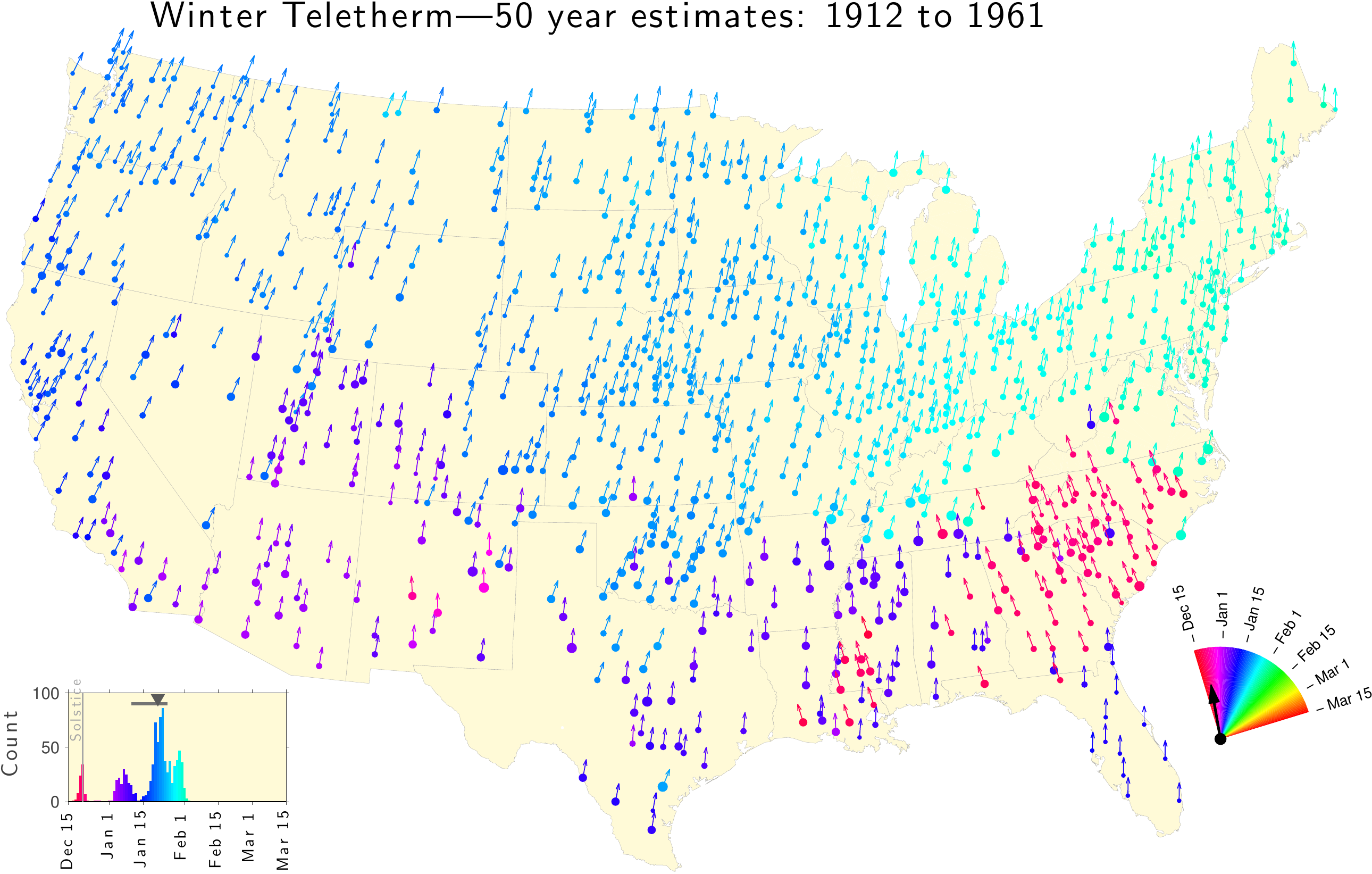}
  \caption{
    Map of the Winter Teletherms and Teletherm Extents estimated for the 50 year
    range 1912--1961,
    to be compared with the equivalent map for 1962--2011 in
    Fig.~\ref{fig:teletherm.universal_teletherm_range_dynamic_wrapped_tmin001_050_1962_to_2011}.
    Fig.~\ref{fig:teletherm_changes001_max_min}B in the main text maps the
    changes
    in Winter Teletherms between these two periods.
    In the bottom left histogram, the gray horizontal line
    shows the interquartile range and the inverted triangle the median.
  }
  \label{fig:teletherm.universal_teletherm_range_dynamic_wrapped_tmin001_050_1912_to_1961}
\end{figure*}

\begin{figure*}[h!]
  \centering
            \includegraphics[width=0.85\textwidth]{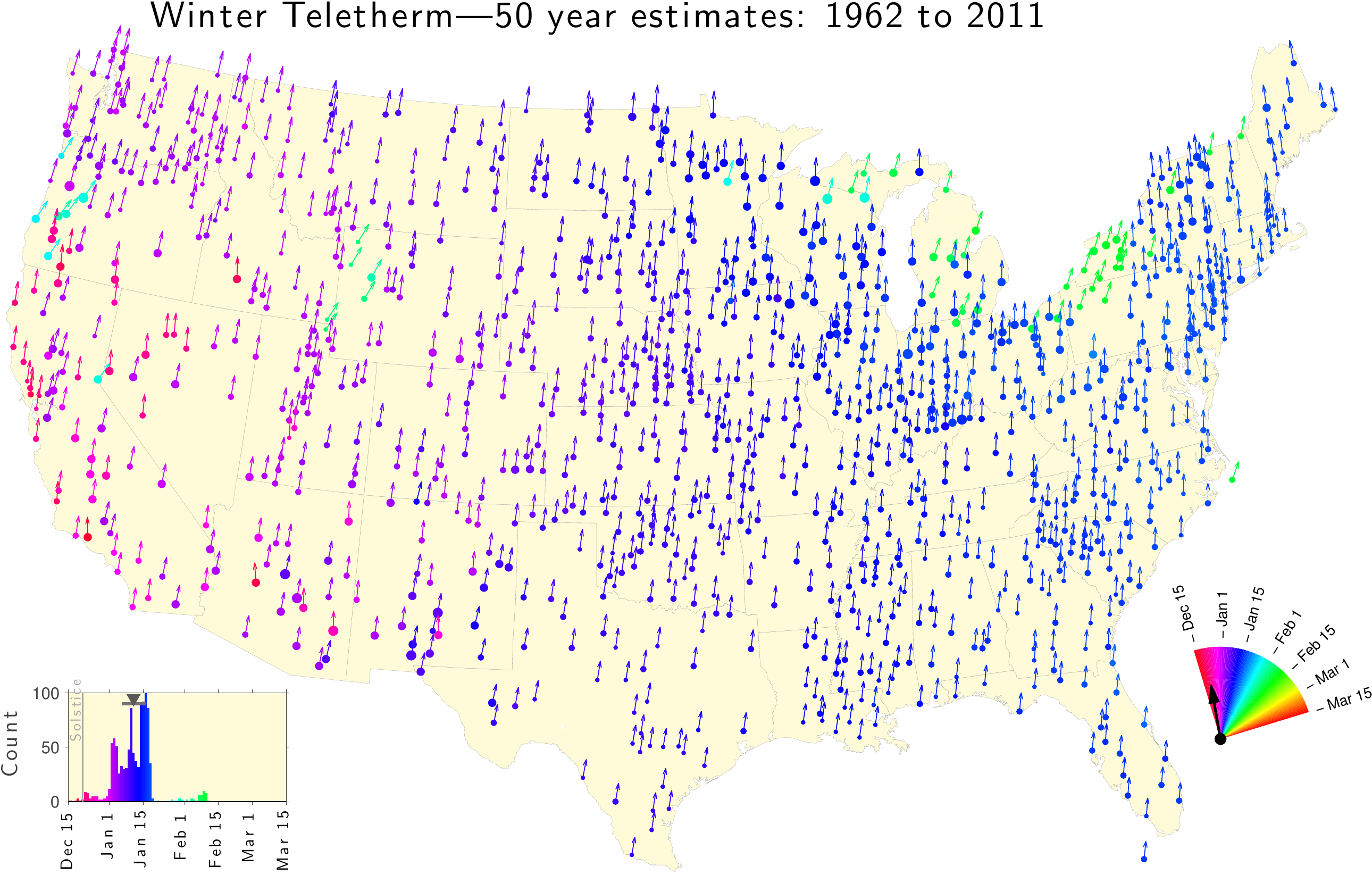}
  \caption{
    Map of the Winter Teletherms and Teletherm Extents estimated for the year
    ranges 1962--2011,
    to be compared with the preceding map in
    Fig.~\ref{fig:teletherm.universal_teletherm_range_dynamic_wrapped_tmin001_050_1912_to_1961}.
    See Fig.~\ref{fig:teletherm_changes001_max_min}B for a map of the changes.
  }
  \label{fig:teletherm.universal_teletherm_range_dynamic_wrapped_tmin001_050_1962_to_2011}
\end{figure*}

\begin{figure*}[h!]
  \centering
      \includegraphics[width=0.85\textwidth]{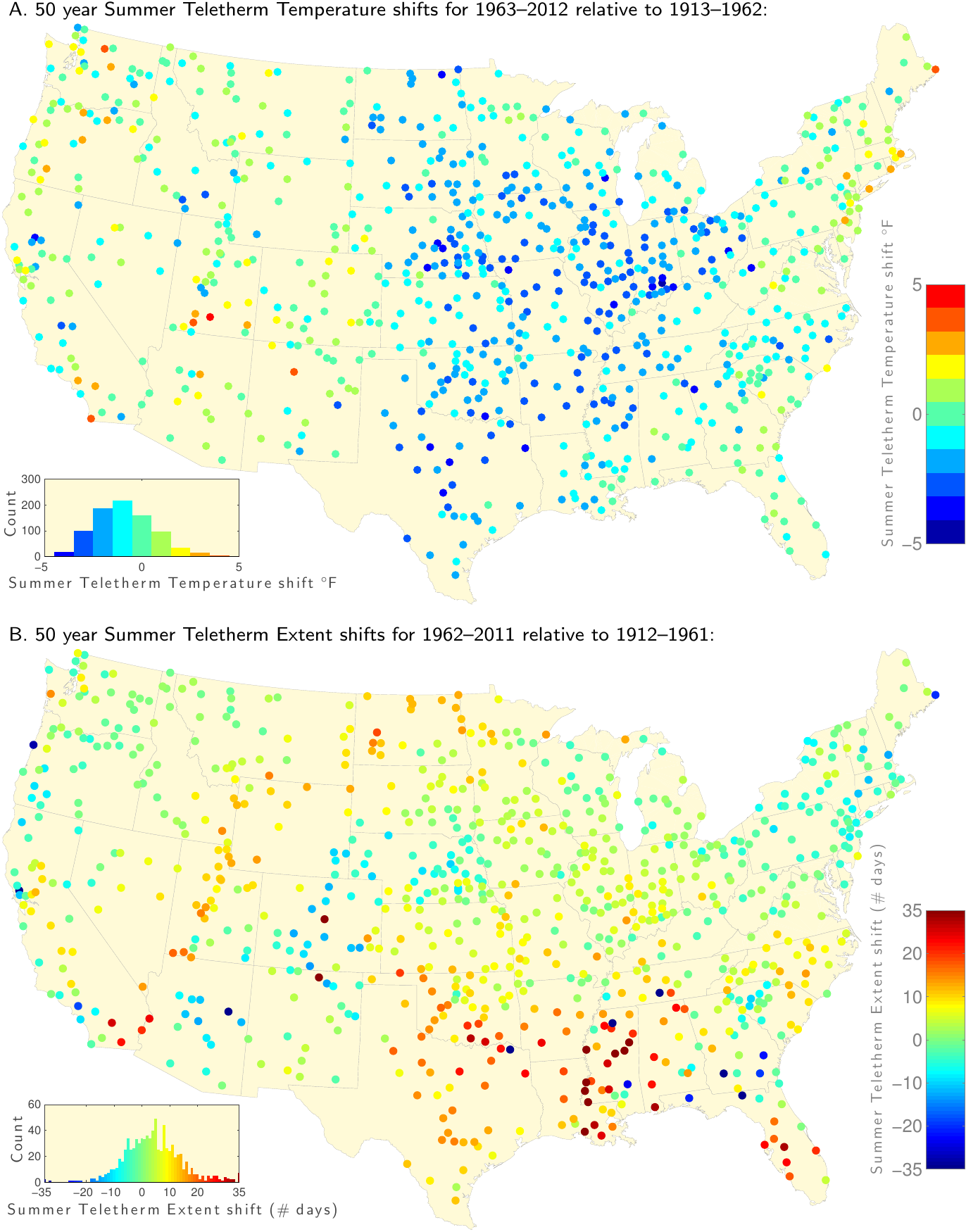}
  \caption{
    Shifts for the Summer Teletherm for 
    \textbf{A:} Temperature
    and
    \textbf{B:} Extent
    derived from
    Figs.~\ref{fig:teletherm.universal_teletherm_range_dynamic_tmax001_050_1913_to_1962}
    and~\ref{fig:teletherm.universal_teletherm_range_dynamic_tmax001_050_1963_to_2012}.
  }
  \label{fig:universal_teletherm_summer_changes_temp_extents}
\end{figure*}

\begin{figure*}[h!]
  \centering
      \includegraphics[width=0.85\textwidth]{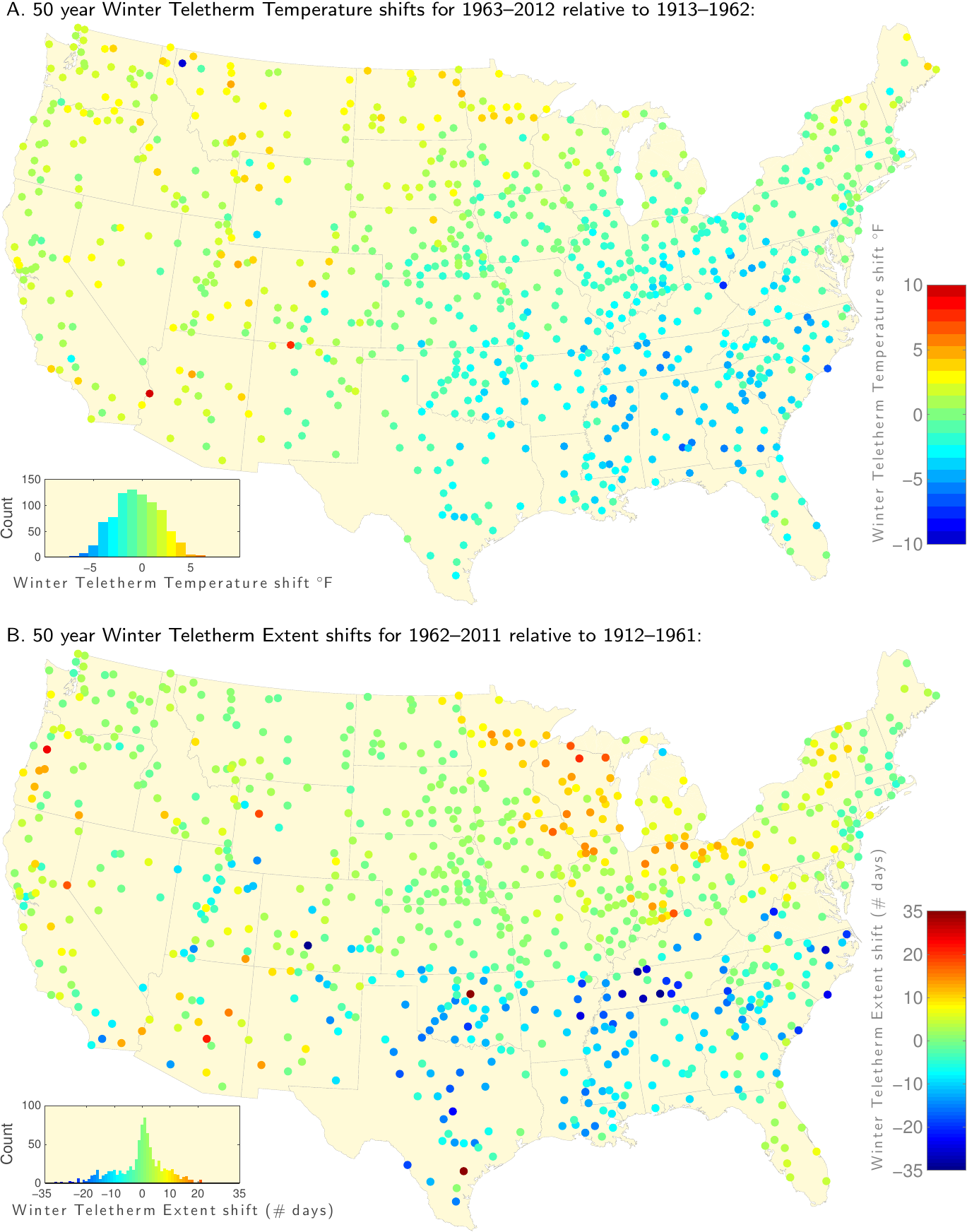}
  \caption{
    Shifts for the Winter Teletherm for 
    \textbf{A:} Temperature
    and
    \textbf{B:} Extent
    derived from
    Figs.~\ref{fig:teletherm.universal_teletherm_range_dynamic_wrapped_tmin001_050_1912_to_1961}
    and~\ref{fig:teletherm.universal_teletherm_range_dynamic_wrapped_tmin001_050_1962_to_2011}.
  }
  \label{fig:universal_teletherm_winter_changes_temp_extents}
\end{figure*}

\begin{figure*}[tp!]
          \centering
      \includegraphics[width=0.85\textwidth]{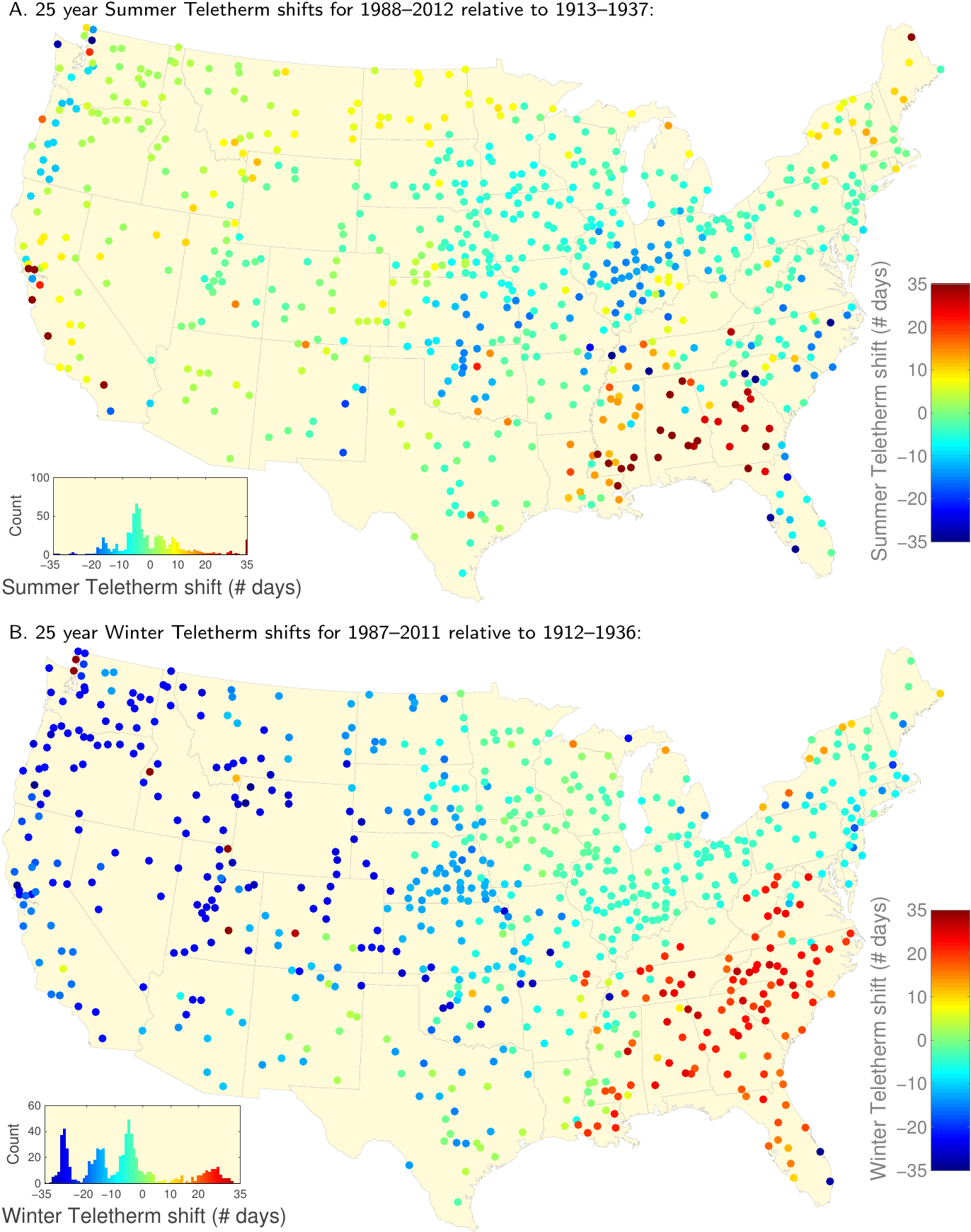}
      \caption{
        Teletherm shifts comparing
        the quarter centuries at the ends of the 1912 to 2012.
        \textbf{A:}
        Summer Teletherm shifts comparing the 25 year periods 1988--2012 relative to
        1912--1937.
        Out of all 1218 stations, 716 (58.8\%)
        have $\ge$ 80\%
        error-free data in both 25 year spans.
        \textbf{B:}
        Winter Teletherm shifts comparing 1987/1988--2011/2012
        relative to 1912/1913--1936/1937.
        A total of 725 out of 1218, 59.5\%, 
        stations have $\ge$ 80\%
        error-free data.
        The overall patterns are consistent with
        those observed for the changes between the consecutive 50 year
        periods spanning the same 100 years, 
        as
        displayed in Fig.~\ref{fig:teletherm_changes001_max_min} in
        the main text.
        For both Teletherms,
        Figs.~\ref{fig:teletherm_changes002_max_min_2},
        \ref{fig:teletherm_changes002_max_min_3},
        and
        \ref{fig:teletherm_changes002_max_min_4}
        show the transitions between consecutive 25 year periods.
      }
      \label{fig:teletherm_changes002_max_min_1}
\end{figure*}

\begin{figure*}[tp!]
          \centering
      \includegraphics[width=0.85\textwidth]{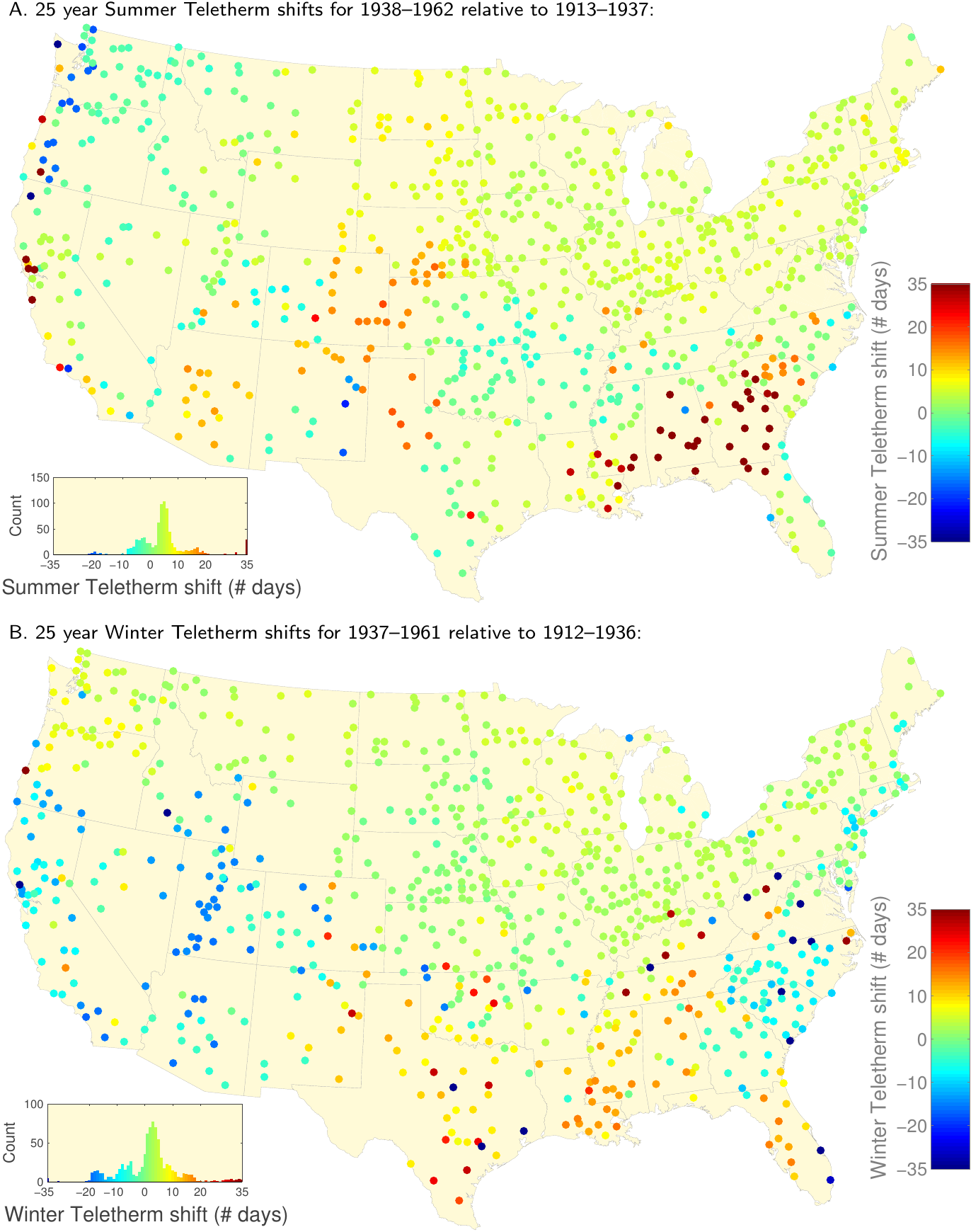}
      \caption{
        Teletherm shifts comparing
        the quarter centuries at the ends of the 1937 to 2012.
        \textbf{A:}
        Summer Teletherm shifts comparing the 25 year period 1938--1962 relative to
        1912--1937
        (837 out of 1218, 68.72\%, stations have acceptable data).
        \textbf{B:}
        Winter Teletherm shifts comparing 1937/1938--1962/1963
        relative to 1912/1913--1936/1937
        (838 out of 1218, 68.80\%, stations have acceptable data).
      }
      \label{fig:teletherm_changes002_max_min_2}
\end{figure*}

\begin{figure*}[tp!]
          \centering
      \includegraphics[width=0.85\textwidth]{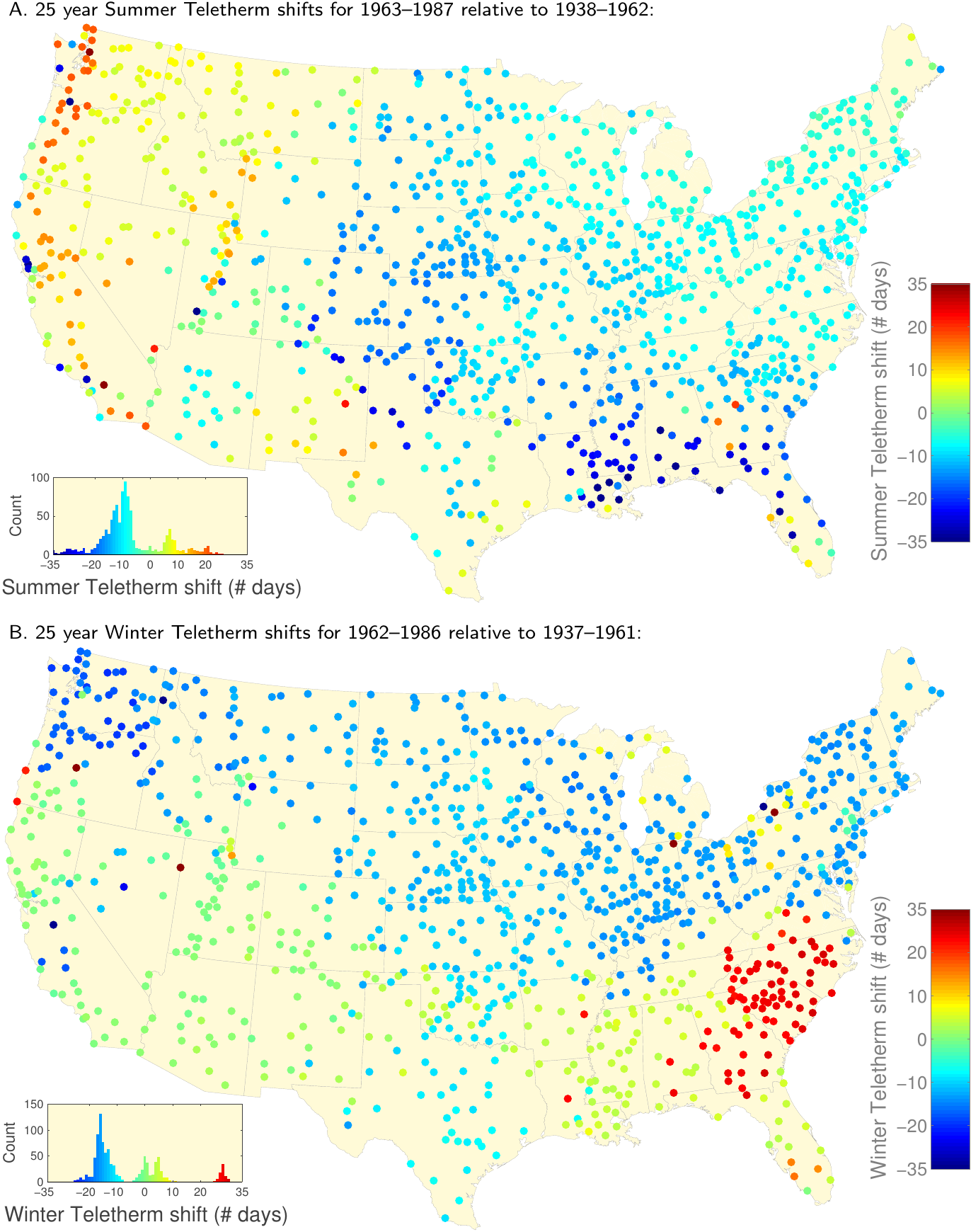}
      \caption{
        Teletherm shifts comparing
        the quarter centuries at the ends of the 1937 to 1987.
        \textbf{A:}
        Summer Teletherm shifts comparing the 25 year period 1963--1987 relative to
        1938--1962
        (1001 out of 1218, 82.18\%, stations have acceptable data).
        \textbf{B:}
        Winter Teletherm shifts comparing 1961/1962--1985/1986
        relative to 1937/1938--1961/1962
        (1000 out of 1218, 82.10\%, stations have acceptable data).
      }
      \label{fig:teletherm_changes002_max_min_3}
\end{figure*}

\begin{figure*}[tp!]
          \centering
      \includegraphics[width=0.85\textwidth]{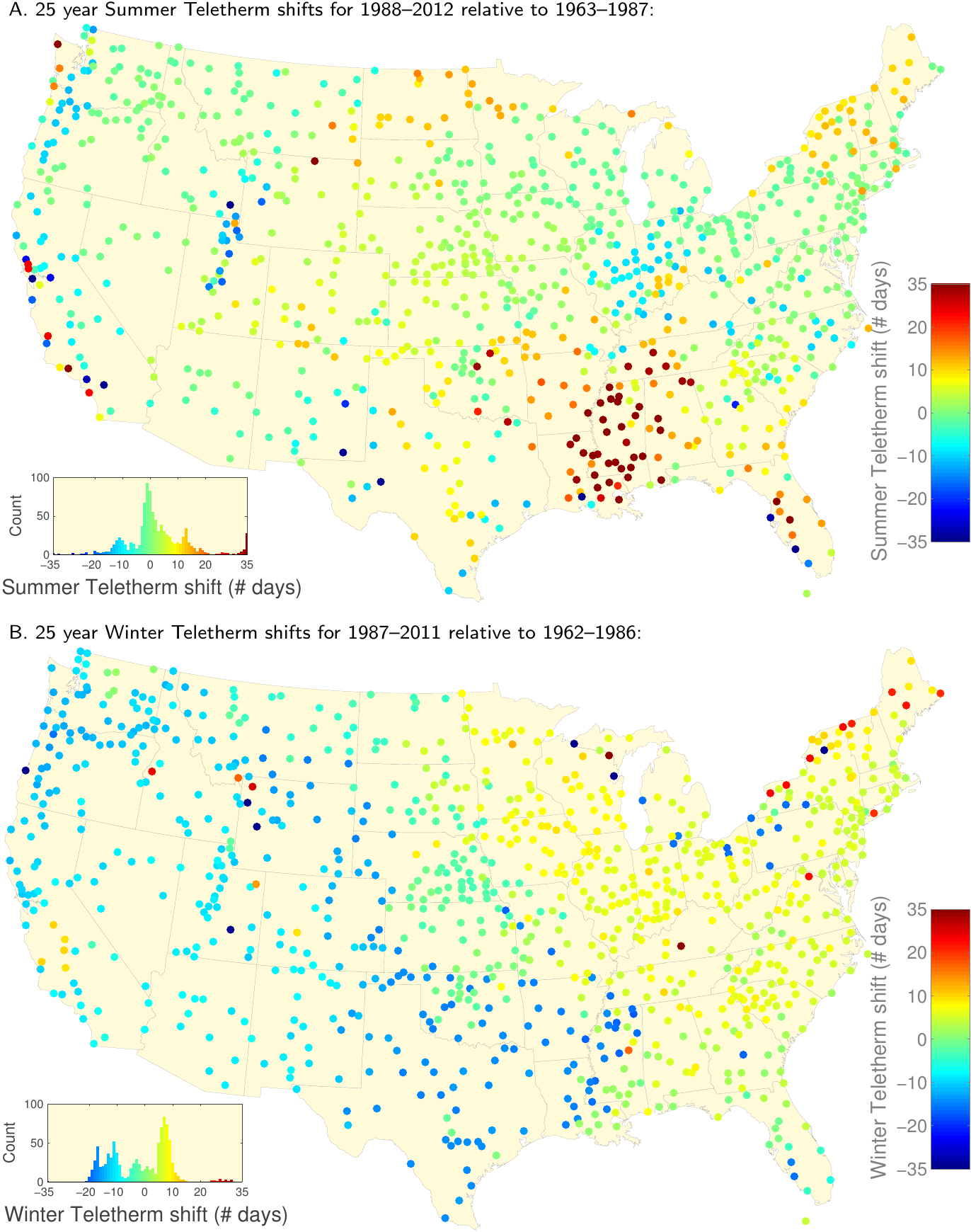}
      \caption{
        Teletherm shifts comparing
        the quarter centuries at the ends of the 1962 to 2012.
        \textbf{A:}
        Summer Teletherm shifts comparing the 25 year period 1988--2012 relative to
        1963--1987
        (941 out of 1218, 77.26\%, stations have acceptable data).
        \textbf{B:}
        Winter Teletherm shifts comparing 1987/1988--2011/2012
        relative to 1962/1963--1986/1987
        (950 out of 1218, 78.00\%, stations have acceptable data).
      }
      \label{fig:teletherm_changes002_max_min_4}
\end{figure*}

\clearpage

\begin{figure*}[tp!]
          \centering
      \includegraphics[width=0.85\textwidth]{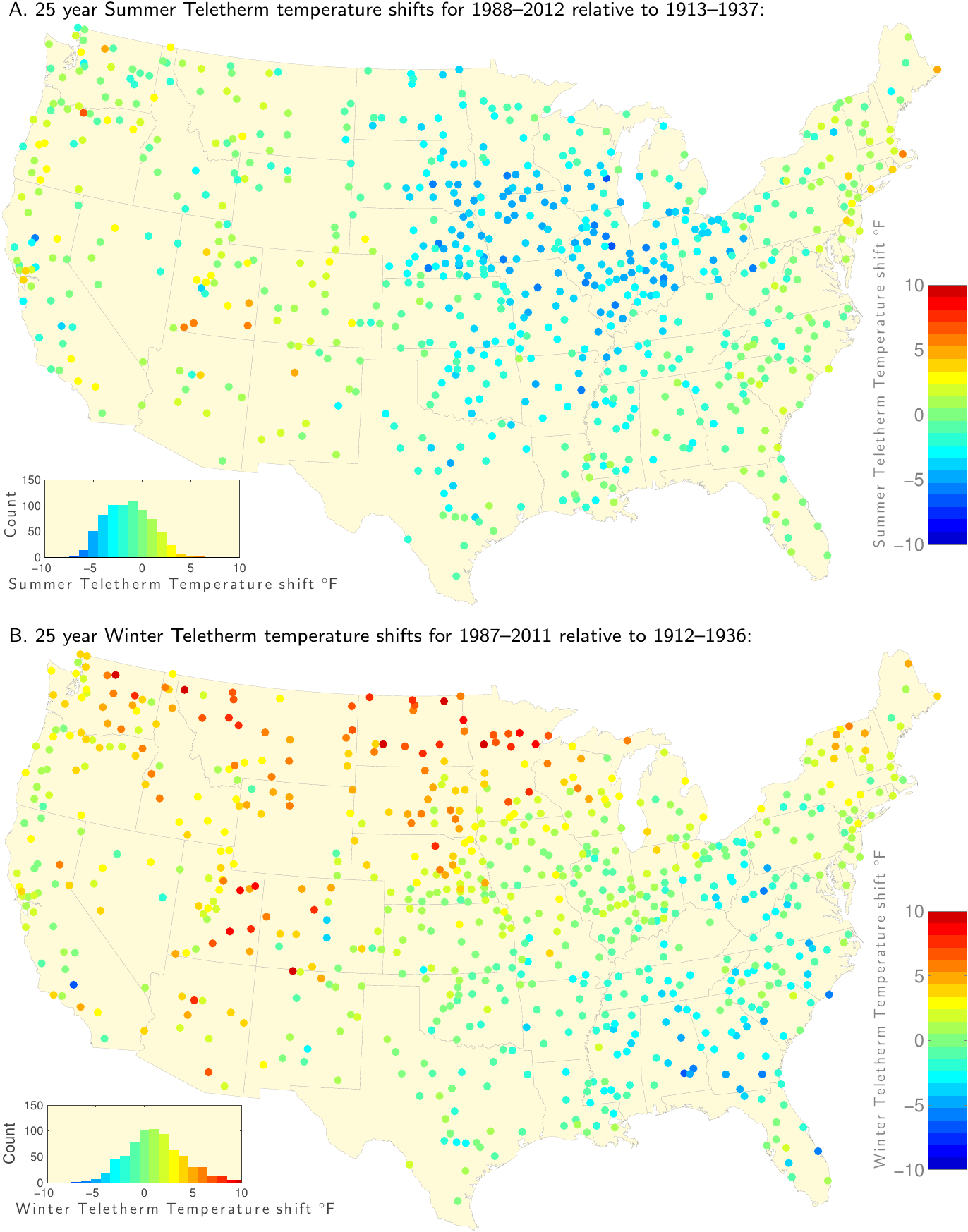}
      \caption{
        Teletherm temperature shifts comparing
        the quarter centuries at the ends of the 1912 to 2012.
        \textbf{A:}
        Summer Teletherm temperature shifts comparing the 25 year periods 1988--2012 relative to
        1912--1937.
        \textbf{B:}
        Winter Teletherm temperature shifts comparing 1987/1988--2011/2012
        relative to 1912/1913--1936/1937.
      }
      \label{fig:teletherm_changes_temperature002_max_min_1}
\end{figure*}

\begin{figure*}[tp!]
          \centering
      \includegraphics[width=0.85\textwidth]{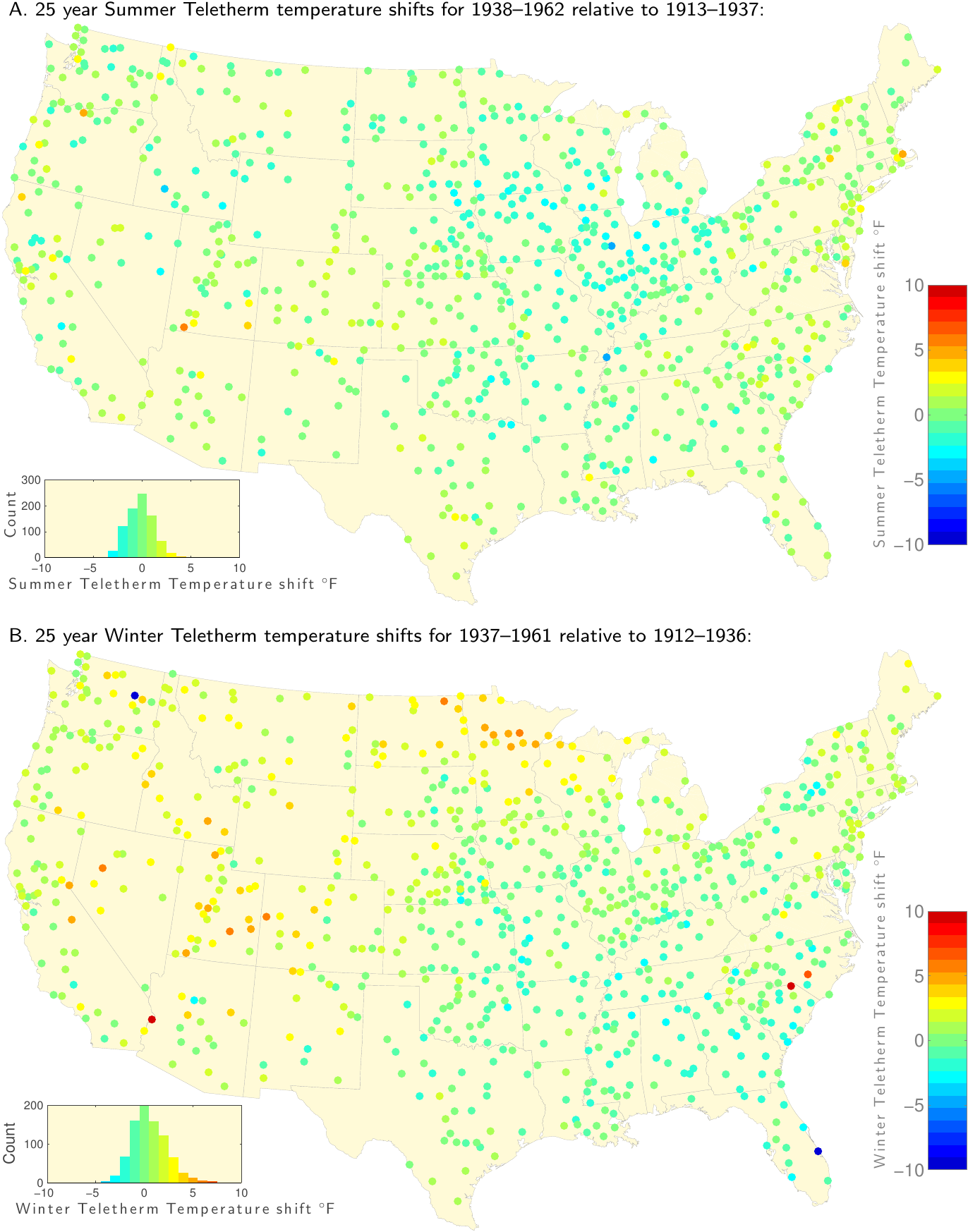}
      \caption{
        Teletherm temperature shifts comparing
        the quarter centuries at the ends of the 1912 to 1963.
        \textbf{A:}
        Summer Teletherm temperature shifts comparing the 25 year period 1938--1962 relative to
        1912--1937.
        \textbf{B:}
        Winter Teletherm temperature shifts comparing 1937/1938--1962/1963
        relative to 1912/1913--1936/1937.
      }
      \label{fig:teletherm_changes_temperature002_max_min_2}
\end{figure*}

\begin{figure*}[tp!]
          \centering
      \includegraphics[width=0.85\textwidth]{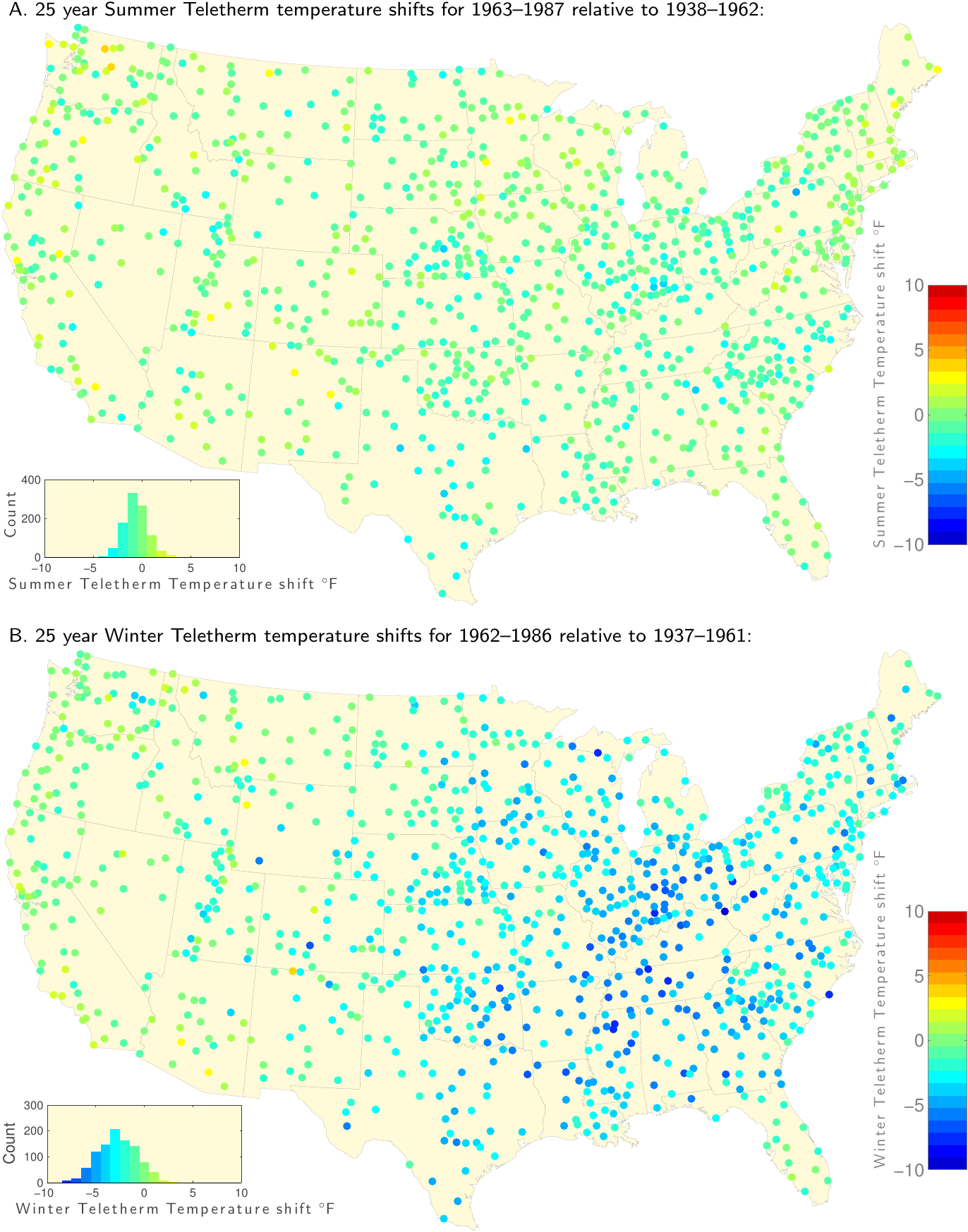}
      \caption{
        Teletherm temperature shifts comparing
        the quarter centuries at the ends of the 1937 to 1987.
        \textbf{A:}
        Summer Teletherm temperature shifts comparing the 25 year period 1963--1987 relative to
        1938--1962.
        \textbf{B:}
        Winter Teletherm temperature shifts comparing 1961/1962--1985/1986
        relative to 1937/1938--1961/1962.
      }
      \label{fig:teletherm_changes_temperature002_max_min_3}
\end{figure*}

\begin{figure*}[tp!]
          \centering
      \includegraphics[width=0.85\textwidth]{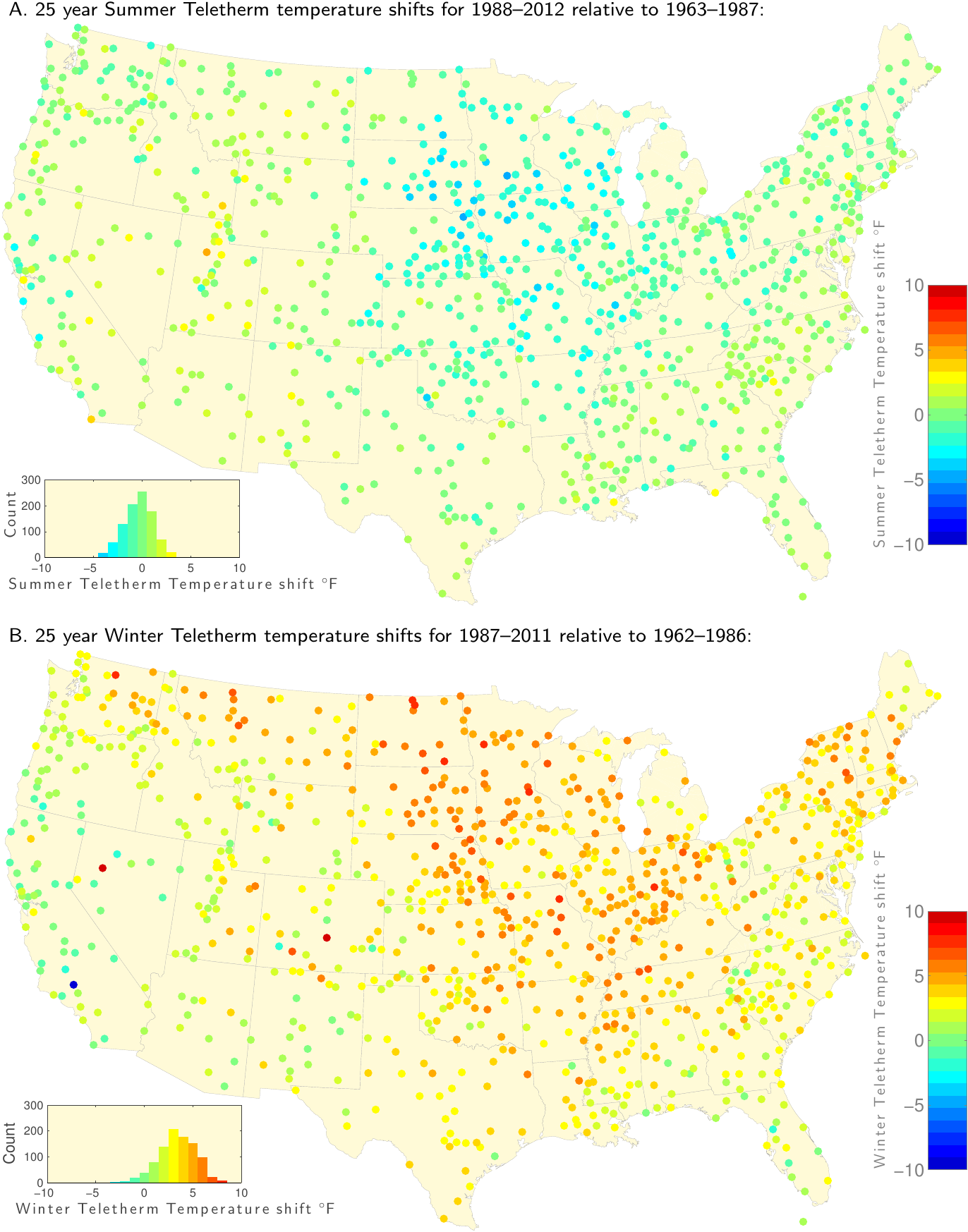}
      \caption{
        Teletherm temperature shifts comparing
        the quarter centuries at the ends of the 1962 to 2012.
        \textbf{A:}
        Summer Teletherm temperature shifts comparing the 25 year period 1988--2012 relative to
        1963--1987.
        \textbf{B:}
        Winter Teletherm temperature shifts comparing 1987/1988--2011/2012
        relative to 1962/1963--1986/1987.
      }
      \label{fig:teletherm_changes_temperature002_max_min_4}
\end{figure*}

\begin{figure*}[tp!]
          \centering
      \includegraphics[width=0.85\textwidth]{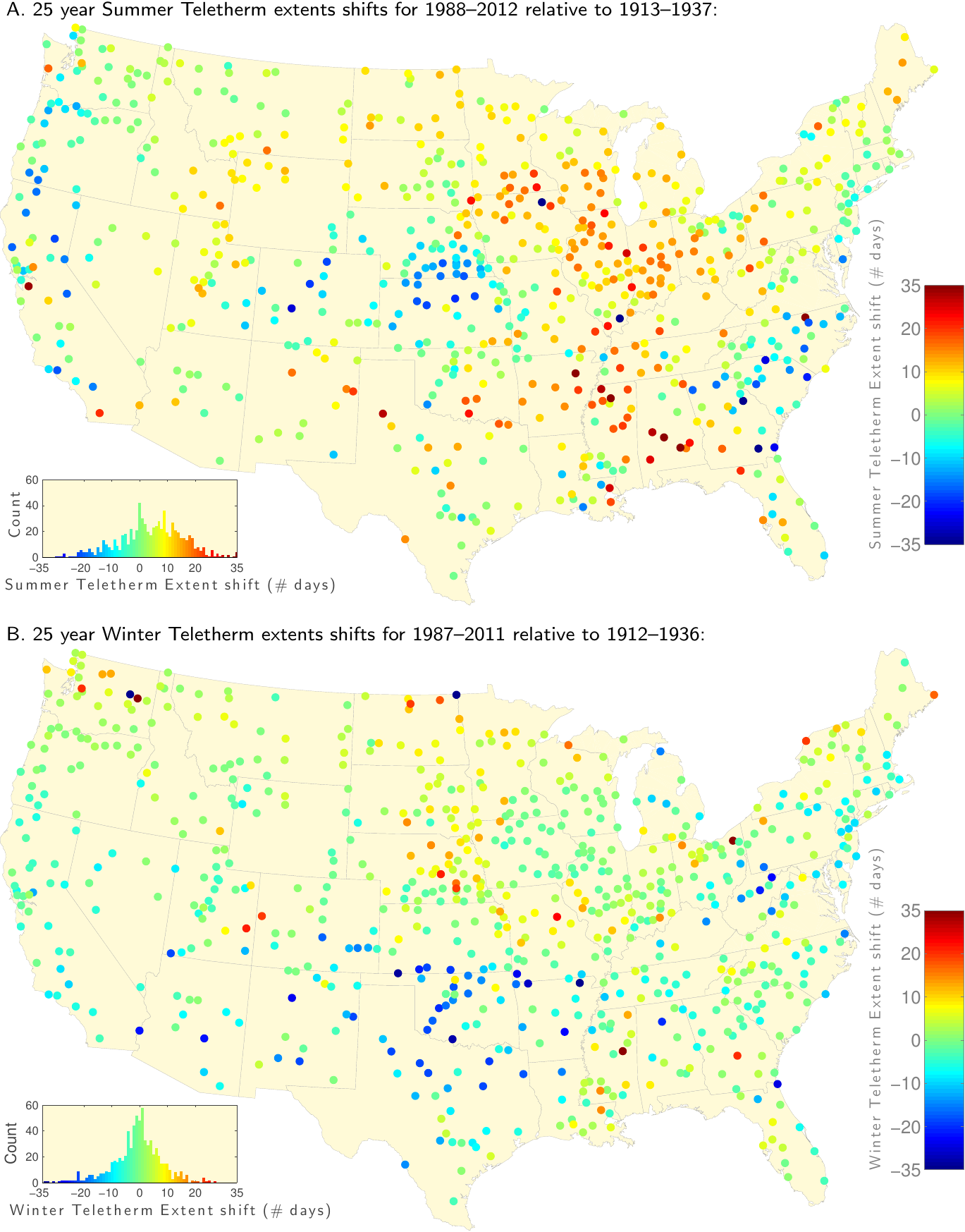}
      \caption{
        Teletherm extent shifts comparing
        the quarter centuries at the ends of the 1912 to 2012.
        \textbf{A:}
        Summer Teletherm extent shifts comparing the 25 year periods 1988--2012 relative to
        1912--1937.
        \textbf{B:}
        Winter Teletherm extent shifts comparing 1987/1988--2011/2012
        relative to 1912/1913--1936/1937.
      }
      \label{fig:teletherm_changes_extents002_max_min_1}
\end{figure*}

\begin{figure*}[tp!]
          \centering
      \includegraphics[width=0.85\textwidth]{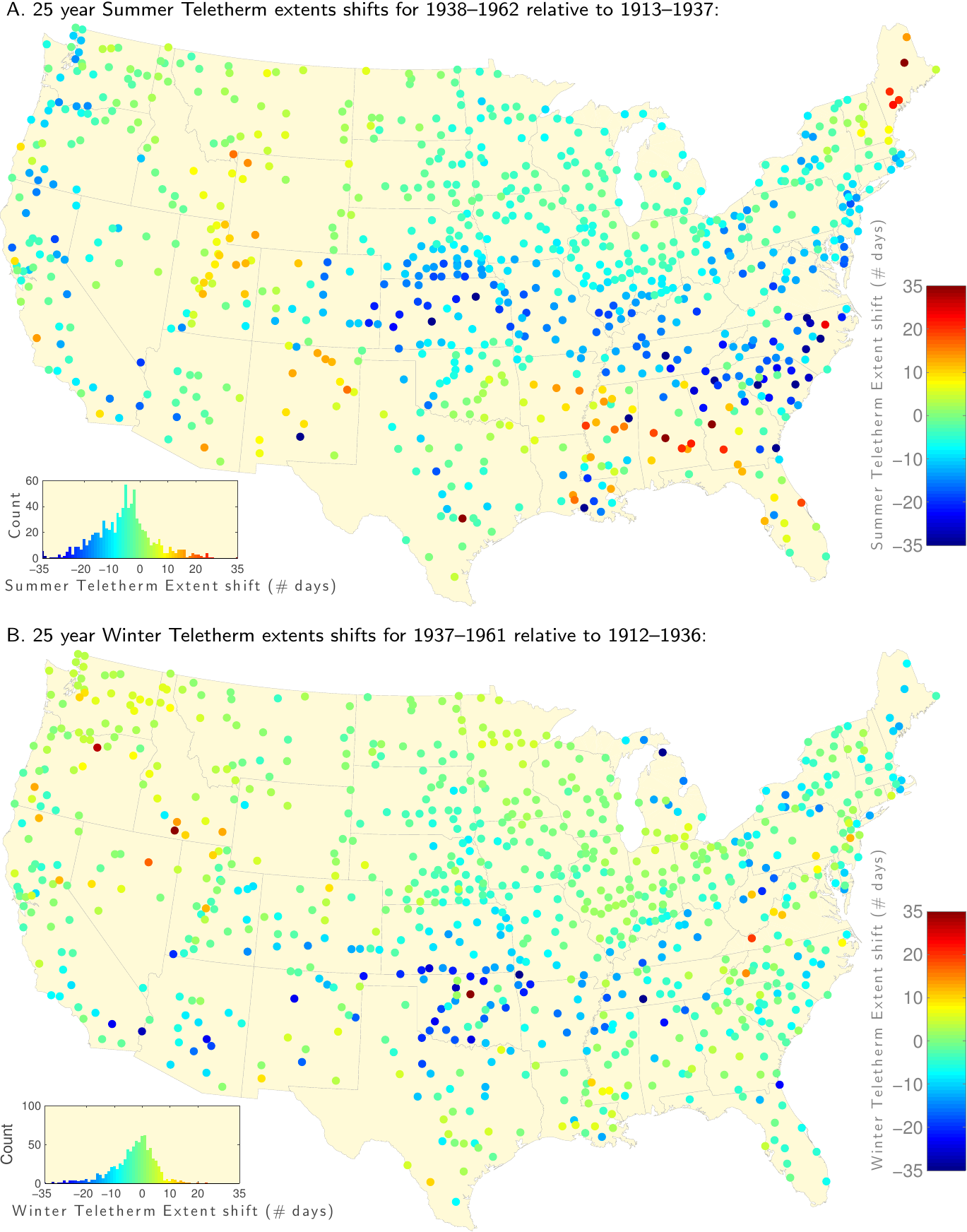}
      \caption{
        Teletherm extent shifts comparing
        the quarter centuries at the ends of the 1912 to 1963.
        \textbf{A:}
        Summer Teletherm extent shifts comparing the 25 year period 1938--1962 relative to
        1912--1937.
        \textbf{B:}
        Winter Teletherm extent shifts comparing 1937/1938--1962/1963
        relative to 1912/1913--1936/1937.
      }
      \label{fig:teletherm_changes_extents002_max_min_2}
\end{figure*}

\begin{figure*}[tp!]
          \centering
      \includegraphics[width=0.85\textwidth]{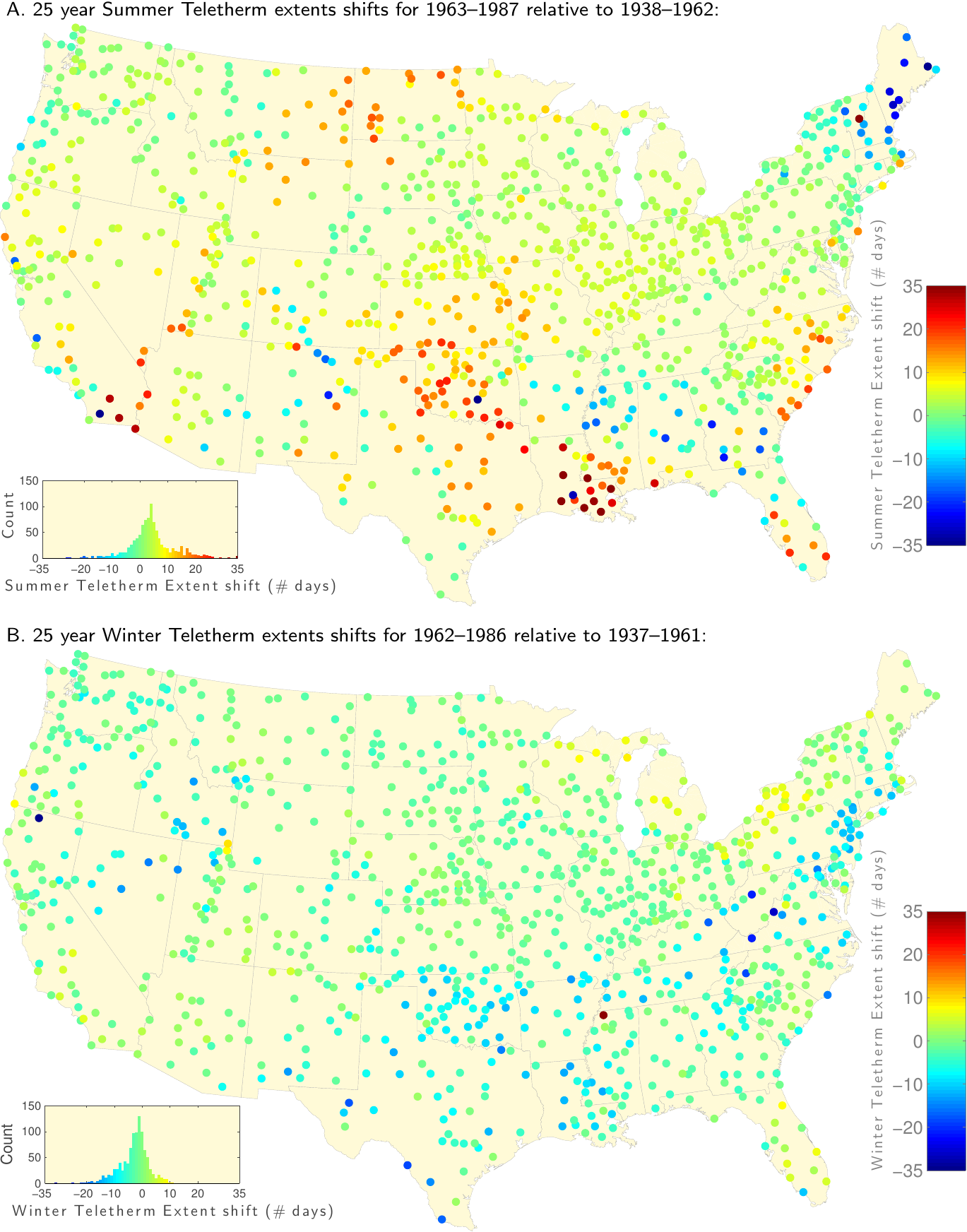}
      \caption{
        Teletherm extent shifts comparing
        the quarter centuries at the ends of the 1937 to 1987.
        \textbf{A:}
        Summer Teletherm extent shifts comparing the 25 year period 1963--1987 relative to
        1938--1962.
        \textbf{B:}
        Winter Teletherm extent shifts comparing 1961/1962--1985/1986
        relative to 1937/1938--1961/1962.
      }
      \label{fig:teletherm_changes_extents002_max_min_3}
\end{figure*}

\begin{figure*}[tp!]
          \centering
      \includegraphics[width=0.85\textwidth]{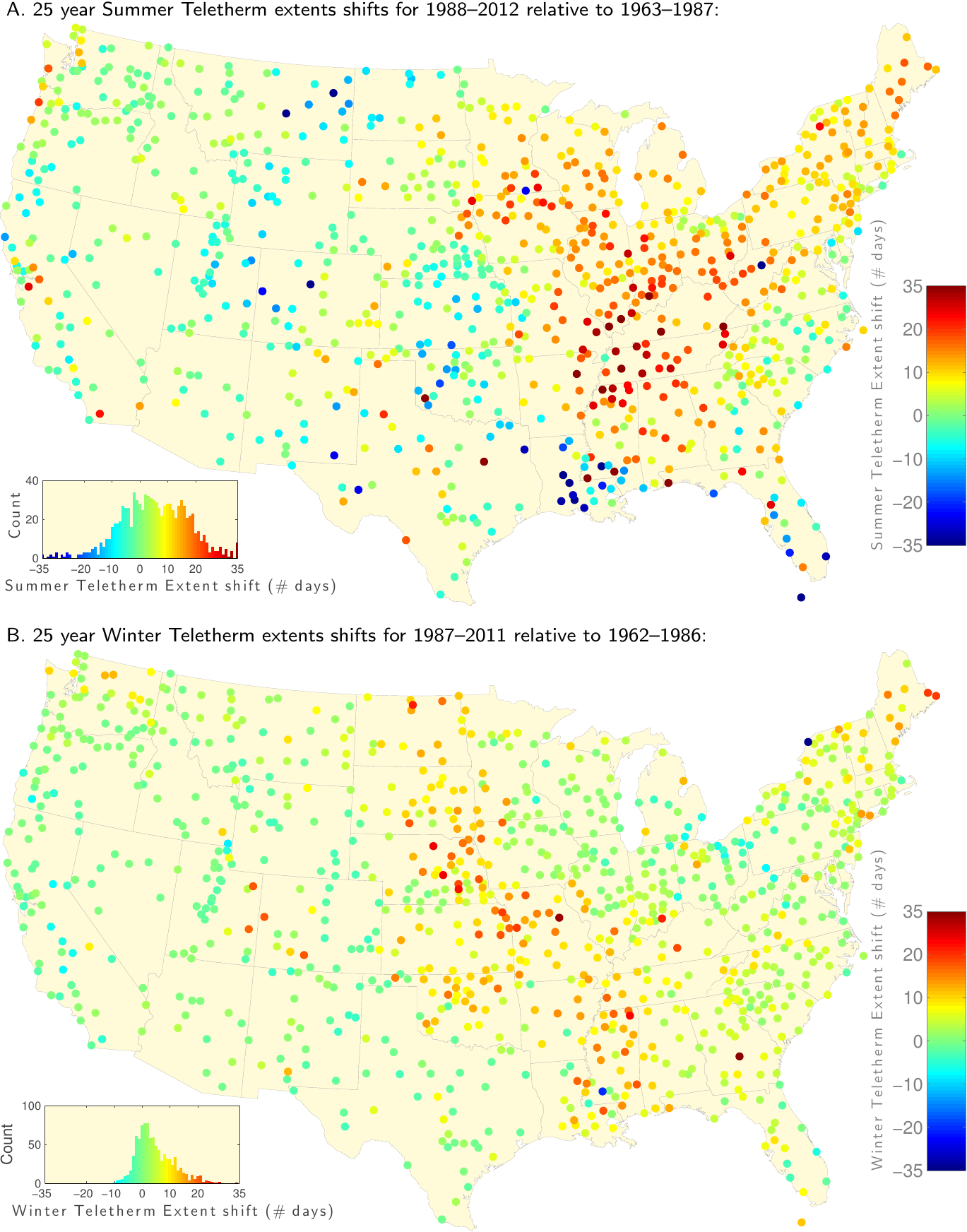}
      \caption{
        Teletherm extent shifts comparing
        the quarter centuries at the ends of the 1962 to 2012.
        \textbf{A:}
        Summer Teletherm extent shifts comparing the 25 year period 1988--2012 relative to
        1963--1987.
        \textbf{B:}
        Winter Teletherm extent shifts comparing 1987/1988--2011/2012
        relative to 1962/1963--1986/1987.
      }
      \label{fig:teletherm_changes_extents002_max_min_4}
\end{figure*}


\begin{thebibliography}{21}%
\makeatletter
\providecommand \@ifxundefined [1]{%
 \@ifx{#1\undefined}
}%
\providecommand \@ifnum [1]{%
 \ifnum #1\expandafter \@firstoftwo
 \else \expandafter \@secondoftwo
 \fi
}%
\providecommand \@ifx [1]{%
 \ifx #1\expandafter \@firstoftwo
 \else \expandafter \@secondoftwo
 \fi
}%
\providecommand \natexlab [1]{#1}%
\providecommand \enquote  [1]{``#1''}%
\providecommand \bibnamefont  [1]{#1}%
\providecommand \bibfnamefont [1]{#1}%
\providecommand \citenamefont [1]{#1}%
\providecommand \href@noop [0]{\@secondoftwo}%
\providecommand \href [0]{\begingroup \@sanitize@url \@href}%
\providecommand \@href[1]{\@@startlink{#1}\@@href}%
\providecommand \@@href[1]{\endgroup#1\@@endlink}%
\providecommand \@sanitize@url [0]{\catcode `\\12\catcode `\$12\catcode
  `\&12\catcode `\#12\catcode `\^12\catcode `\_12\catcode `\%12\relax}%
\providecommand \@@startlink[1]{}%
\providecommand \@@endlink[0]{}%
\providecommand \url  [0]{\begingroup\@sanitize@url \@url }%
\providecommand \@url [1]{\endgroup\@href {#1}{\urlprefix }}%
\providecommand \urlprefix  [0]{URL }%
\providecommand \Eprint [0]{\href }%
\providecommand \doibase [0]{http://dx.doi.org/}%
\providecommand \selectlanguage [0]{\@gobble}%
\providecommand \bibinfo  [0]{\@secondoftwo}%
\providecommand \bibfield  [0]{\@secondoftwo}%
\providecommand \translation [1]{[#1]}%
\providecommand \BibitemOpen [0]{}%
\providecommand \bibitemStop [0]{}%
\providecommand \bibitemNoStop [0]{.\EOS\space}%
\providecommand \EOS [0]{\spacefactor3000\relax}%
\providecommand \BibitemShut  [1]{\csname bibitem#1\endcsname}%
\let\auto@bib@innerbib\@empty
\bibitem [{\citenamefont {Arguez}\ \emph {et~al.}(2012)\citenamefont {Arguez},
  \citenamefont {Durre}, \citenamefont {Applequist}, \citenamefont {Vose},
  \citenamefont {Squires}, \citenamefont {Yin}, \citenamefont {Heim},\ and\
  \citenamefont {Owen}}]{arguez2012a}%
  \BibitemOpen
  \bibfield  {author} {\bibinfo {author} {\bibfnamefont {A.}~\bibnamefont
  {Arguez}}, \bibinfo {author} {\bibfnamefont {I.}~\bibnamefont {Durre}},
  \bibinfo {author} {\bibfnamefont {S.}~\bibnamefont {Applequist}}, \bibinfo
  {author} {\bibfnamefont {R.~S.}\ \bibnamefont {Vose}}, \bibinfo {author}
  {\bibfnamefont {M.~F.}\ \bibnamefont {Squires}}, \bibinfo {author}
  {\bibfnamefont {X.}~\bibnamefont {Yin}}, \bibinfo {author} {\bibfnamefont
  {R.~R.}\ \bibnamefont {Heim}, \bibfnamefont {Jr.}}, \ and\ \bibinfo {author}
  {\bibfnamefont {T.~W.}\ \bibnamefont {Owen}},\ }\href@noop {} {\bibfield
  {journal} {\bibinfo  {journal} {Bull. Amer. Meteor. Soc.}\ }\textbf {\bibinfo
  {volume} {93}},\ \bibinfo {pages} {1687} (\bibinfo {year}
  {2012})}\BibitemShut {NoStop}%
\bibitem [{\citenamefont {Stainforth}\ \emph {et~al.}(2005)\citenamefont
  {Stainforth}, \citenamefont {Aina}, \citenamefont {Christensen},
  \citenamefont {Collins}, \citenamefont {Faull}, \citenamefont {Frame},
  \citenamefont {Kettleborough}, \citenamefont {Knight}, \citenamefont
  {Martin}, \citenamefont {Murphy} \emph {et~al.}}]{stainforth2005a}%
  \BibitemOpen
  \bibfield  {author} {\bibinfo {author} {\bibfnamefont {D.~A.}\ \bibnamefont
  {Stainforth}}, \bibinfo {author} {\bibfnamefont {T.}~\bibnamefont {Aina}},
  \bibinfo {author} {\bibfnamefont {C.}~\bibnamefont {Christensen}}, \bibinfo
  {author} {\bibfnamefont {M.}~\bibnamefont {Collins}}, \bibinfo {author}
  {\bibfnamefont {N.}~\bibnamefont {Faull}}, \bibinfo {author} {\bibfnamefont
  {D.~J.}\ \bibnamefont {Frame}}, \bibinfo {author} {\bibfnamefont {J.~A.}\
  \bibnamefont {Kettleborough}}, \bibinfo {author} {\bibfnamefont
  {S.}~\bibnamefont {Knight}}, \bibinfo {author} {\bibfnamefont
  {A.}~\bibnamefont {Martin}}, \bibinfo {author} {\bibfnamefont
  {J.}~\bibnamefont {Murphy}},  \emph {et~al.},\ }\href@noop {} {\bibfield
  {journal} {\bibinfo  {journal} {Nature}\ }\textbf {\bibinfo {volume} {433}},\
  \bibinfo {pages} {403} (\bibinfo {year} {2005})}\BibitemShut {NoStop}%
\bibitem [{\citenamefont {Karl}\ \emph {et~al.}(2015)\citenamefont {Karl},
  \citenamefont {Arguez}, \citenamefont {Huang}, \citenamefont {Lawrimore},
  \citenamefont {McMahon}, \citenamefont {Menne}, \citenamefont {Peterson},
  \citenamefont {Vose},\ and\ \citenamefont {Zhang}}]{karl2015a}%
  \BibitemOpen
  \bibfield  {author} {\bibinfo {author} {\bibfnamefont {T.~R.}\ \bibnamefont
  {Karl}}, \bibinfo {author} {\bibfnamefont {A.}~\bibnamefont {Arguez}},
  \bibinfo {author} {\bibfnamefont {B.}~\bibnamefont {Huang}}, \bibinfo
  {author} {\bibfnamefont {J.~H.}\ \bibnamefont {Lawrimore}}, \bibinfo {author}
  {\bibfnamefont {J.~R.}\ \bibnamefont {McMahon}}, \bibinfo {author}
  {\bibfnamefont {M.~J.}\ \bibnamefont {Menne}}, \bibinfo {author}
  {\bibfnamefont {T.~C.}\ \bibnamefont {Peterson}}, \bibinfo {author}
  {\bibfnamefont {R.~S.}\ \bibnamefont {Vose}}, \ and\ \bibinfo {author}
  {\bibfnamefont {H.-M.}\ \bibnamefont {Zhang}},\ }\href@noop {} {\bibfield
  {journal} {\bibinfo  {journal} {Science Magazine}\ }\textbf {\bibinfo
  {volume} {348}},\ \bibinfo {pages} {1469} (\bibinfo {year}
  {2015})}\BibitemShut {NoStop}%
\bibitem [{\citenamefont {Barkemeyer}\ \emph {et~al.}(2015)\citenamefont
  {Barkemeyer}, \citenamefont {Dessai}, \citenamefont {Monge-Sanz},
  \citenamefont {Renzi},\ and\ \citenamefont {Napolitano}}]{barkemeyer2015a}%
  \BibitemOpen
  \bibfield  {author} {\bibinfo {author} {\bibfnamefont {R.}~\bibnamefont
  {Barkemeyer}}, \bibinfo {author} {\bibfnamefont {S.}~\bibnamefont {Dessai}},
  \bibinfo {author} {\bibfnamefont {B.}~\bibnamefont {Monge-Sanz}}, \bibinfo
  {author} {\bibfnamefont {B.~G.}\ \bibnamefont {Renzi}}, \ and\ \bibinfo
  {author} {\bibfnamefont {G.}~\bibnamefont {Napolitano}},\ }\href@noop {}
  {\bibfield  {journal} {\bibinfo  {journal} {Nature Climate Change}\ ,\
  \bibinfo {pages} {10.1038/nclimate2824}} (\bibinfo {year}
  {2015})}\BibitemShut {NoStop}%
\bibitem [{\citenamefont {et~al.}(2015)}]{oreilly2015a}%
  \BibitemOpen
  \bibfield  {author} {\bibinfo {author} {\bibfnamefont {C.~M.~O.}\
  \bibnamefont {et~al.}},\ }\href@noop {} {\bibfield  {journal} {\bibinfo
  {journal} {Geophysical Research Letters}\ }\textbf {\bibinfo {volume} {42}},\
  \bibinfo {pages} {10,773} (\bibinfo {year} {2015})}\BibitemShut {NoStop}%
\bibitem [{\citenamefont {Antilla}(2005)}]{antilla2005a}%
  \BibitemOpen
  \bibfield  {author} {\bibinfo {author} {\bibfnamefont {L.}~\bibnamefont
  {Antilla}},\ }\href@noop {} {\bibfield  {journal} {\bibinfo  {journal}
  {Global Environmental Change}\ }\textbf {\bibinfo {volume} {15}},\ \bibinfo
  {pages} {338} (\bibinfo {year} {2005})}\BibitemShut {NoStop}%
\bibitem [{\citenamefont {Lempert}(2015)}]{lempert2015a}%
  \BibitemOpen
  \bibfield  {author} {\bibinfo {author} {\bibfnamefont {R.~J.}\ \bibnamefont
  {Lempert}},\ }\href@noop {} {\bibfield  {journal} {\bibinfo  {journal}
  {Nature Climate Change}\ }\textbf {\bibinfo {volume} {5}},\ \bibinfo {pages}
  {914} (\bibinfo {year} {2015})}\BibitemShut {NoStop}%
\bibitem [{\citenamefont {Mann}\ and\ \citenamefont {Park}(1996)}]{mann1996a}%
  \BibitemOpen
  \bibfield  {author} {\bibinfo {author} {\bibfnamefont {M.}~\bibnamefont
  {Mann}}\ and\ \bibinfo {author} {\bibfnamefont {J.}~\bibnamefont {Park}},\
  }\href@noop {} {\bibfield  {journal} {\bibinfo  {journal} {Geophys. Res.
  Lett.}\ }\textbf {\bibinfo {volume} {23}},\ \bibinfo {pages} {1111} (\bibinfo
  {year} {1996})}\BibitemShut {NoStop}%
\bibitem [{\citenamefont {Sparks}\ and\ \citenamefont
  {Menzel}(2002)}]{sparks2002a}%
  \BibitemOpen
  \bibfield  {author} {\bibinfo {author} {\bibfnamefont {T.~H.}\ \bibnamefont
  {Sparks}}\ and\ \bibinfo {author} {\bibfnamefont {A.}~\bibnamefont
  {Menzel}},\ }\href@noop {} {\bibfield  {journal} {\bibinfo  {journal}
  {International Journal of Climatology}\ }\textbf {\bibinfo {volume} {22}},\
  \bibinfo {pages} {1715} (\bibinfo {year} {2002})}\BibitemShut {NoStop}%
\bibitem [{\citenamefont {Schwartz}\ \emph {et~al.}(2006)\citenamefont
  {Schwartz}, \citenamefont {Ahas},\ and\ \citenamefont
  {Aasa}}]{schwartz2006a}%
  \BibitemOpen
  \bibfield  {author} {\bibinfo {author} {\bibfnamefont {M.~D.}\ \bibnamefont
  {Schwartz}}, \bibinfo {author} {\bibfnamefont {R.}~\bibnamefont {Ahas}}, \
  and\ \bibinfo {author} {\bibfnamefont {A.}~\bibnamefont {Aasa}},\ }\href@noop
  {} {\bibfield  {journal} {\bibinfo  {journal} {Global Change Biology}\
  }\textbf {\bibinfo {volume} {12}},\ \bibinfo {pages} {343} (\bibinfo {year}
  {2006})}\BibitemShut {NoStop}%
\bibitem [{\citenamefont {Stine}\ \emph {et~al.}(2009)\citenamefont {Stine},
  \citenamefont {Huybers},\ and\ \citenamefont {Y.}}]{stine2009a}%
  \BibitemOpen
  \bibfield  {author} {\bibinfo {author} {\bibfnamefont {A.~R.}\ \bibnamefont
  {Stine}}, \bibinfo {author} {\bibfnamefont {P.}~\bibnamefont {Huybers}}, \
  and\ \bibinfo {author} {\bibfnamefont {F.~I.}\ \bibnamefont {Y.}},\
  }\href@noop {} {\bibfield  {journal} {\bibinfo  {journal} {Nature}\ }\textbf
  {\bibinfo {volume} {457}},\ \bibinfo {pages} {435} (\bibinfo {year}
  {2009})}\BibitemShut {NoStop}%
\bibitem [{\citenamefont {Betts}(2011)}]{betts2011a}%
  \BibitemOpen
  \bibfield  {author} {\bibinfo {author} {\bibfnamefont {A.~K.}\ \bibnamefont
  {Betts}},\ }\href@noop {} {\bibfield  {journal} {\bibinfo  {journal}
  {Weather}\ }\textbf {\bibinfo {volume} {66}},\ \bibinfo {pages} {245}
  (\bibinfo {year} {2011})}\BibitemShut {NoStop}%
\bibitem [{\citenamefont {Lai}(2016)}]{lai2016a}%
  \BibitemOpen
  \bibfield  {author} {\bibinfo {author} {\bibfnamefont {K.~K.~R.}\
  \bibnamefont {Lai}},\ }\href@noop {} {\enquote {\bibinfo {title} {How much
  warmer was your city in 2015?}}\ } (\bibinfo {year} {2016}),\ \bibinfo {note}
  {{T}he {N}ew {Y}ork {T}imes's interactive weather chart:
  \url{http://www.nytimes.com/interactive/2016/02/19/us/2015-year-in-weather-temperature-precipitation.html};
  accessed February 20, 2016}\BibitemShut {NoStop}%
\bibitem [{\citenamefont {Menne}\ \emph {et~al.}(2012)\citenamefont {Menne},
  \citenamefont {Durre}, \citenamefont {Vose}, \citenamefont {Gleason},\ and\
  \citenamefont {Houston}}]{menne2012a}%
  \BibitemOpen
  \bibfield  {author} {\bibinfo {author} {\bibfnamefont {M.~J.}\ \bibnamefont
  {Menne}}, \bibinfo {author} {\bibfnamefont {I.}~\bibnamefont {Durre}},
  \bibinfo {author} {\bibfnamefont {R.~S.}\ \bibnamefont {Vose}}, \bibinfo
  {author} {\bibfnamefont {B.~E.}\ \bibnamefont {Gleason}}, \ and\ \bibinfo
  {author} {\bibfnamefont {T.~G.}\ \bibnamefont {Houston}},\ }\href@noop {}
  {\bibfield  {journal} {\bibinfo  {journal} {Journal of Atmospheric and
  Oceanic Technology}\ }\textbf {\bibinfo {volume} {29}},\ \bibinfo {pages}
  {897} (\bibinfo {year} {2012})}\BibitemShut {NoStop}%
\bibitem [{\citenamefont {Trenberth}(2015)}]{trenberth2015a}%
  \BibitemOpen
  \bibfield  {author} {\bibinfo {author} {\bibfnamefont {K.~E.}\ \bibnamefont
  {Trenberth}},\ }\href@noop {} {\bibfield  {journal} {\bibinfo  {journal}
  {Science Magazine}\ }\textbf {\bibinfo {volume} {349}},\ \bibinfo {pages}
  {691} (\bibinfo {year} {2015})}\BibitemShut {NoStop}%
\bibitem [{\citenamefont {Collins}\ \emph {et~al.}(2006)\citenamefont {Collins}
  \emph {et~al.}}]{collins2006a}%
  \BibitemOpen
  \bibfield  {author} {\bibinfo {author} {\bibfnamefont {W.~D.}\ \bibnamefont
  {Collins}} \emph {et~al.},\ }\href@noop {} {\bibfield  {journal} {\bibinfo
  {journal} {J. Climate}\ }\textbf {\bibinfo {volume} {19}},\ \bibinfo {pages}
  {2122} (\bibinfo {year} {2006})}\BibitemShut {NoStop}%
\bibitem [{\citenamefont {Mearns}\ \emph {et~al.}(2009)\citenamefont {Mearns},
  \citenamefont {Gutowski}, \citenamefont {Jones}, \citenamefont {Leung},
  \citenamefont {McGinnis}, \citenamefont {Nunes},\ and\ \citenamefont
  {Qian}}]{mearns2009a}%
  \BibitemOpen
  \bibfield  {author} {\bibinfo {author} {\bibfnamefont {L.~O.}\ \bibnamefont
  {Mearns}}, \bibinfo {author} {\bibfnamefont {W.~J.}\ \bibnamefont
  {Gutowski}}, \bibinfo {author} {\bibfnamefont {R.}~\bibnamefont {Jones}},
  \bibinfo {author} {\bibfnamefont {L.-Y.}\ \bibnamefont {Leung}}, \bibinfo
  {author} {\bibfnamefont {S.}~\bibnamefont {McGinnis}}, \bibinfo {author}
  {\bibfnamefont {A.~M.~B.}\ \bibnamefont {Nunes}}, \ and\ \bibinfo {author}
  {\bibfnamefont {Y.}~\bibnamefont {Qian}},\ }\href@noop {} {\bibfield
  {journal} {\bibinfo  {journal} {EOS}\ }\textbf {\bibinfo {volume} {90}},\
  \bibinfo {pages} {311} (\bibinfo {year} {2009})}\BibitemShut {NoStop}%
\bibitem [{\citenamefont {Mearns}\ \emph {et~al.}(2014)\citenamefont {Mearns}
  \emph {et~al.}}]{mearns2014a}%
  \BibitemOpen
  \bibfield  {author} {\bibinfo {author} {\bibfnamefont {L.~O.}\ \bibnamefont
  {Mearns}} \emph {et~al.},\ }\href@noop {} {\enquote {\bibinfo {title} {The
  {N}orth {A}merican {R}egional {C}limate {C}hange {A}ssessment {P}rogram
  dataset, {N}ational {C}enter for {A}tmospheric {R}esearch {E}arth {S}ystem
  {G}rid data portal, {B}oulder, {C}{O}},}\ } (\bibinfo {year} {2014}),\
  \bibinfo {note} {data downloaded 2102-03-28.
  doi:10.5065/D6RN35ST}\BibitemShut {NoStop}%
\bibitem [{\citenamefont {Mearns}\ \emph {et~al.}(2012)\citenamefont {Mearns}
  \emph {et~al.}}]{mearns2012a}%
  \BibitemOpen
  \bibfield  {author} {\bibinfo {author} {\bibfnamefont {L.~O.}\ \bibnamefont
  {Mearns}} \emph {et~al.},\ }\href@noop {} {\bibfield  {journal} {\bibinfo
  {journal} {Bull. Amer. Meteor. Soc.}\ }\textbf {\bibinfo {volume} {93}},\
  \bibinfo {pages} {1337} (\bibinfo {year} {2012})}\BibitemShut {NoStop}%
\bibitem [{\citenamefont {Greasby}\ and\ \citenamefont
  {Sain}(2012)}]{greasby2012a}%
  \BibitemOpen
  \bibfield  {author} {\bibinfo {author} {\bibfnamefont {T.~A.}\ \bibnamefont
  {Greasby}}\ and\ \bibinfo {author} {\bibfnamefont {S.~R.}\ \bibnamefont
  {Sain}},\ }\href@noop {} {\enquote {\bibinfo {title} {Assessing uncertainty
  in climate model ensembles via annual temperature profiles},}\ } (\bibinfo
  {year} {2012}),\ \bibinfo {note} {{U}npublished manuscript}\BibitemShut
  {NoStop}%
\bibitem [{\citenamefont {Greasby}\ and\ \citenamefont
  {Sain}(2011)}]{greasby2011a}%
  \BibitemOpen
  \bibfield  {author} {\bibinfo {author} {\bibfnamefont {T.}~\bibnamefont
  {Greasby}}\ and\ \bibinfo {author} {\bibfnamefont {S.}~\bibnamefont {Sain}},\
  }\href@noop {} {\bibfield  {journal} {\bibinfo  {journal} {Journal of
  Agricultural, Biological, and Environmental Statistics}\ }\textbf {\bibinfo
  {volume} {16}},\ \bibinfo {pages} {571} (\bibinfo {year} {2011})}\BibitemShut
  {NoStop}%
\end{thebibliography}
\end{document}